\documentclass[pra,letterpaper,aps,10pt,superscriptaddress,floatfix,showpacs]{revtex4-1}
\usepackage{graphicx}
\usepackage{amsmath}
\usepackage{autobreak}
\usepackage{amsfonts}
\usepackage{amssymb}
\usepackage{epsfig}
\usepackage[pdftex]{color}
\usepackage{amsmath,graphicx,amssymb,braket,xcolor,subfigure,upgreek}
\usepackage[colorlinks, linkcolor=blue, citecolor=blue, urlcolor=blue, breaklinks=true]{hyperref}
\usepackage{microtype}
\usepackage{bbm}
\usepackage{color}
\usepackage{mathtools}
\usepackage{rotating}

\newcommand{\fref}[1]{Fig. \ref{#1}}

\begin{document}

\title{Continuous narrowband lasing with coherently driven V-level atoms}
\author{Christoph Hotter}
\affiliation{Institut f\"ur Theoretische Physik, Universit\"at Innsbruck, Technikerstr. 21a, A-6020 Innsbruck, Austria}
\author{David Plankensteiner}
\affiliation{Institut f\"ur Theoretische Physik, Universit\"at Innsbruck, Technikerstr. 21a, A-6020 Innsbruck, Austria}
\author{Helmut Ritsch}
\affiliation{Institut f\"ur Theoretische Physik, Universit\"at Innsbruck, Technikerstr. 21a, A-6020 Innsbruck, Austria}
\date{\today}

\begin{abstract}
Simultaneous strong coherent pumping of the two transitions of a V-level atom with very different decay rates has been predicted to create almost perfect inversion on the narrower transition. Using the example of the blue and red transitions in Strontium we show that for suitable operating conditions the corresponding resonant gain can be used to continuously operate a laser on the narrow transition. In particular, for a strong detuning of the pump field with respect to the narrow transition, coherent laser emission occurs close to the bare atomic transition frequency exhibiting only a negligible contribution from coherent pump light scattered into the lasing mode. Calculations of the cavity output spectrum show that the resulting laser linewidth can get much smaller than the bandwidth of the pump light and even the natural linewidth of the narrow atomic transition. Its frequency is closely tied to the atomic transition frequency for properly chosen atom numbers. Simulations including atomic motion show Doppler cooling on the strong transition with minor motion heating on the lasing transition, so that continuous laser operation in the presence of a magneto-optical trap should be possible with current experimental technology.
\end{abstract}


\maketitle

\section{Introduction}
It has been a longstanding dream of AMO physics to implement an active optical frequency standard by operating a continuous laser on a narrow atomic transition in close analogy to microwave masers~\cite{schawlow1958infrared,strelnitski1995hydrogen,goldenberg1960atomic}. This new class of lasers would exhibit superb accuracy, precision, and robustness against thermal noise~\cite{Haake1993superradiant,meiser2009prospects}. In particular, in view of the recent development and outstanding success of optical atomic clocks, worldwide efforts towards implementations of such an active optical clock have enormously grown in the past few years~\cite{norcia2016cold, norcia2018frequency, laske2019pulse, schaffer2020lasing, Gogyan2020characterisation}. This was further fueled by the prospects of superior stability and accuracy theoretically predicted for superradiant clock lasers \cite{meiser2009prospects,bohnet2012steady,maier2014superradiant,norcia2016superradiance}. Since in such bad-cavity lasers the phase coherence of the system is stored in the atomic gain medium, the laser setup is largely insensitive to technical fluctuations for an isolated atomic gas. In particular, thermal fluctuations of the cavity mirrors~\cite{numata2004thermal} are strongly suppressed. An important challenge that hinders the further development in this direction lies in achieving the necessary steady-state inversion of such a narrow transition: the implementation of efficient and minimally perturbative pump schemes proves extremely difficult. Thus, finding a suitable driving mechanism constitutes a central issue for the realization of highly stable active optical frequency standards~\cite{Chen2009active}.

One possible route create a continuously inverted intra-cavity gain medium is to send a sufficiently dense beam of excited atoms through a cavity~\cite{Kazakov2014active, Chen2019continuous}. In this process a $\pi$-pulse is applied to the atoms just before they enter the lasing region. However, such a setup requires a cold and dense atomic beam which has to be perfectly controlled. Furthermore, the coherence can only be stored within the part of the atomic medium that is within the active lasing region, which requires a high intra-cavity atom number. Therefore, finding a mechanism to create steady-state inversion by repumping atoms within the cavity is highly desirable since this would allow straightforward continuous operation.

\begin{figure}[hb]
\includegraphics[width=0.9\linewidth]{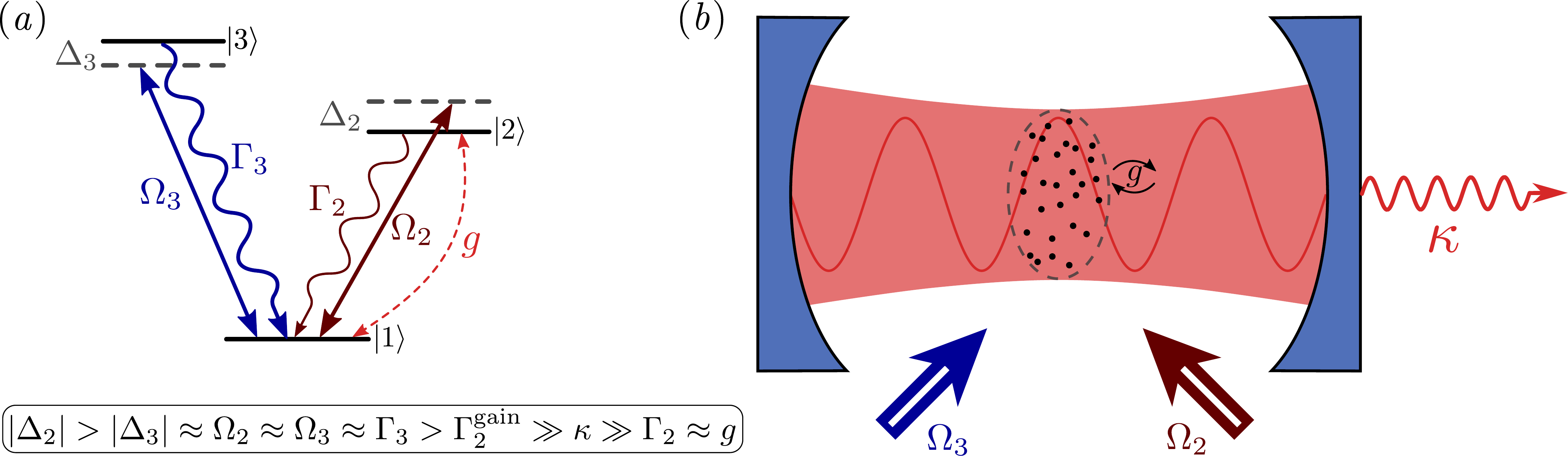}
\caption{\emph{Continuous lasing of a driving V-type atom}. In the sketch (a) we show the considered coherently pumped V-level system. Population inversion can be achieved on the narrower transition, which in our case is the transition $| 1 \rangle \leftrightarrow | 2 \rangle$ with a negative detuning on the broad transition, $\Delta_3 <0$, and $\Delta_2>0$. (b) An ensemble of such atoms is then placed inside a cavity and the narrow transition is coupled to the cavity mode with coupling strength $g$. For later reference we show the frequency hierarchy in the parameter region needed for lasing action in (a). Note that $\Gamma_2^\text{gain}$ denotes the power-broadened transition $| 1 \rangle \leftrightarrow | 2 \rangle$ which provides the gain for lasing.}
\label{fig:schematic}
\end{figure}

Unfortunately, inversion on a transition cannot simply be achieved by coherent pumping as stimulated emission always compensates absorption. However, in Ref.~\cite{Meduri1993dynamical} a surprising mechanism leading to steady-state population inversion on the narrower transition of a V-type atom via coherent driving was shown theoretically. In the model considered therein, both transitions of the V-level atom are driven coherently and no direct decay channel between the two excited states is present, as depicted in \fref{fig:schematic}(a). To the best of our knowledge, this scheme was so far not shown experimentally. Interestingly, lasing using a V-level system was recently observed experimentally~\cite{Gothe2019Continuous}. While the pump mechanism for this virtual-state lasing appears very similar at first sight, it turns out that the system is operated in a distinctly different parameter regime, which leads to anti-Stokes Raman gain with no inversion on the narrow transition. 

While the appearance of inversion on the narrow transition has already been theoretically shown, the usefulness of this unconventional driving scheme in a lasing setup [see \fref{fig:schematic}(b)] remained an open question. The aim of this paper is to address precisely this point: we start by reviewing the driving scheme. Then, we show that steady-state inversion can still be achieved for pump lasers with a realistic spectral linewidth far above the natural linewidth of the narrow atomic transition. We proceed by coupling the inverted, narrow transition to an optical cavity and show that the system behaves like a laser with a clear threshold for sufficient gain with increasing atom number. As an important feature we predict that the spectral linewidth of the output laser light can be well below that of the narrow transition and the pump light. Note, however, that even though the linewidth of the cavity we consider is much larger than the natural linewidth of the narrow transition, we do not operate in the typical low intensity bad cavity regime: due to the power broadening induced by the strong driving laser the effective linewidth of the gain medium is much larger than that of the optical resonator. Thus, the system resembles more a conventional laser than a superradiant one.

Another important aspect for stable operation of our laser is the thermal back-action of lasing on the gain medium: this includes heating due to optical pumping or photon recoil from spontaneous emission, which causes line broadening as well as particle loss via heating in the gain medium. Hence, we provide an estimate of these effects for the considered unconventional driving scheme. We show that Doppler cooling from the two pump lasers occurs. Therefore, for operating parameters that allow lasing (even though not optimal ones), we find that the kinetic energy of an atomic ensemble subjected to the driving scheme is limited to the Doppler temperature of the broad transition, which should allow stationary operation. 


\section{Steady-State Population Inversion via Coherent Driving} \label{sec:pump_mech}
In this section, we provide a brief review of the driving scheme from Ref.~\cite{Meduri1993dynamical}. Moreover, we investigate the influence of a finite pump laser linewidth and the time scale of the process. Note, that the key point is that population inversion can be achieved without a direct irreversible process that causes gain in the excited state. Rather, the scheme here is based on an indirect incoherent process~\cite{Meduri1993dynamical}.

\begin{figure*}
\begin{minipage}[b]{0.3532\linewidth}
\centering
\includegraphics[width=\textwidth]{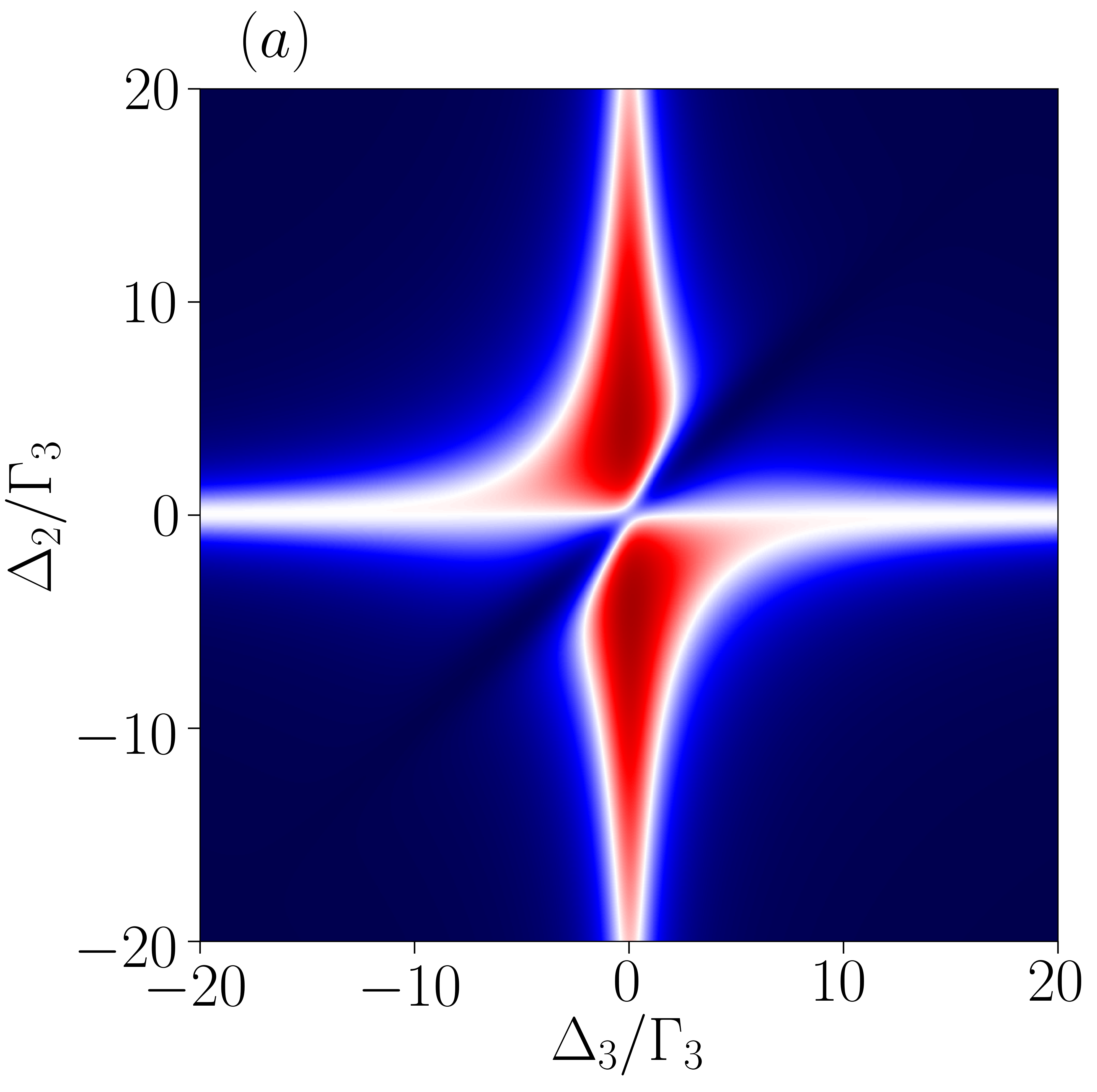}
\end{minipage}
\hspace{1cm}
\begin{minipage}[b]{0.426\linewidth}
\centering
\includegraphics[width=\textwidth]{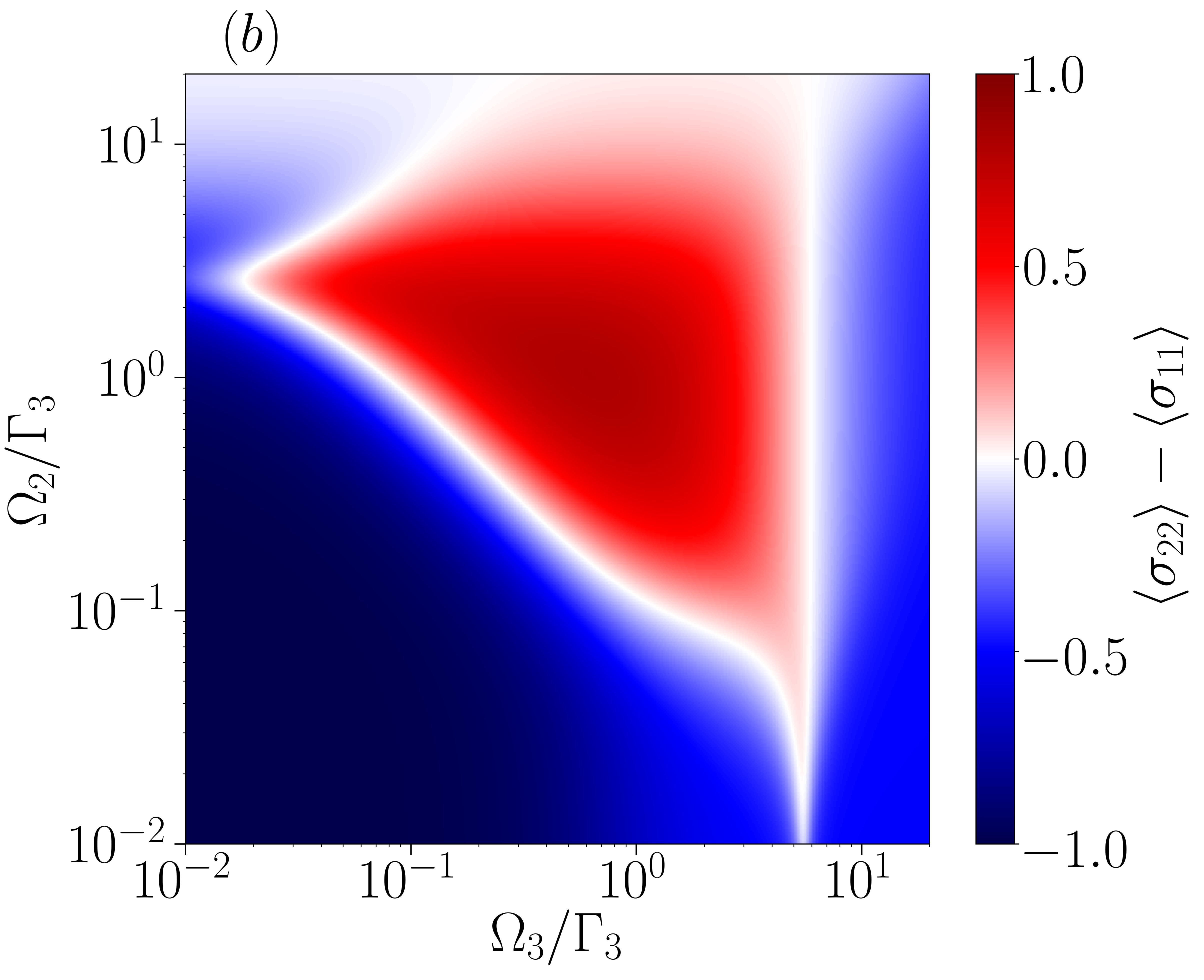}
\end{minipage}
\caption{\emph{Steady-state inversion of a V-level atom}. The figures show the population difference of the narrow transition $\ket{1}\leftrightarrow\ket{2}$ as a function of both driving laser detunings (a) and Rabi frequencies (b), respectively. Red areas indicate population inversion. The parameters when kept constant are $\Delta_2 = 5\Gamma_3$, $\Delta_3 = -1 \Gamma_3$, $\Omega_2 = 0.5 \Gamma_3$ and $\Omega_3 = 0.5 \Gamma_3$.}
\label{fig:v-level_inversion}
\end{figure*}

We consider a V-level atom, which is coherently pumped on both transitions, as depicted in \fref{fig:schematic}(a). The ground state is denoted by $| 1 \rangle$ and the two excited states by $| 2 \rangle$ and $| 3 \rangle$. Decay from an excited state $| i \rangle$ to the ground state $| 1 \rangle$ occurs at a rate $\Gamma_i$. Each transition $| 1 \rangle \leftrightarrow | i \rangle$ is driven coherently with the respective Rabi frequency $\Omega_i$. The difference of the driving laser frequency $\omega_{\ell i}$ and the atomic resonance frequency of a transition $\omega_i$ is given by the detuning $\Delta_i = \omega_{\ell i} - \omega_i$. The time evolution of the density matrix $\rho$ for this system is described by the master equation
\begin{equation}
\dot{\rho} = - i [H, \rho] + \mathcal{L}[\rho].
\label{eq:master_eq}
\end{equation}
In the rotating frame of both pump lasers the Hamiltonian reads
\begin{equation}
H = \sum_{ i=\{2,3\}} -\Delta_i \sigma_{ii} + \Omega_i (\sigma_{i1} + \sigma_{1i})
\label{eq:Hamiltonian}
\end{equation}
with the atomic operators defined by $\sigma_{ij}  \coloneqq | i \rangle \langle j |$.
The dissipative processes are accounted for by the Liouvillian in standard Lindblad form. For the decay from an excited state $| i \rangle$ to the ground state $| 1\rangle$ the Liouvillian term reads
\begin{equation}
\mathcal{L}_\mathrm{\Gamma}[\rho] = \sum_{i = \{ 2,3 \}} \frac{\Gamma_i }{2} (2 \sigma_{1i} \rho \sigma_{i1} - \sigma_{ii} \rho - \rho \sigma_{ii})
\label{eq:liouvillian_decay}
\end{equation}
Let us stress here again that there is no decay channel from $| 3 \rangle$ to $| 2 \rangle$ or vice versa.

A necessary property for the V-level atom to be able to exhibit steady-state population inversion with this scheme, is that the atom needs to have a big ratio between the two decay rates. In our case we choose the transition $| 1 \rangle \leftrightarrow | 2 \rangle$ to be the narrower one, i.e. $\Gamma_3/\Gamma_2 \gg 1$. For example, in the case of ${}^{88}\text{Sr}$ we get a ratio of $\Gamma_3/\Gamma_2 \approx 4266$ for the transitions $\ket{1}\equiv {}^1 \text{S}_0$, $\ket{2}\equiv {}^1 \text{P}_1$, and $\ket{3}\equiv {}^3\text{P}_1$, with the corresponding decay rates $\Gamma_2 = 2 \pi 7.5 \text{kHz}$ and $\Gamma_3 = 2 \pi 32 \text{MHz}$, respectively. For ${}^{174}\text{Yb}$ the ratio for the same transitions is approximately $\Gamma_3/\Gamma_2 \approx 160$. This is still sufficient to create population inversion, but leads to a lower maximal steady-state population inversion compared to ${}^{88}\text{Sr}$. In general, a larger ratio of the decay rates leads to a higher maximal population inversion. Note, that through this pumping scheme steady-state population inversion can only be achieved on the narrower transition. In the following calculations, we will always use the case of ${}^{88}\text{Sr}$, i.e. with a decay rate ratio of $\Gamma_3/\Gamma_2 = 4266$.

Using the above Hamiltonian \eqref{eq:Hamiltonian} and Liouvillian \eqref{eq:liouvillian_decay}, we compute the steady state of the system. \fref{fig:v-level_inversion} shows scans of the population difference $\langle \sigma_{22} \rangle - \langle \sigma_{11} \rangle$ over tunable system parameters, namely both detunings and Rabi frequencies. We can see that it is possible to achieve an inversion of almost $100\%$ for ${}^{88}\text{Sr}$. Note, that all parameters are in units of $\Gamma_3$. Hence, a relatively large Rabi frequency on the transition $| 1 \rangle \leftrightarrow | 2 \rangle$ ($\Omega_2 \gtrsim 0.1\Gamma_3$) is needed to achieve a significant population inversion.

Investigating the inversion when varying the detunings [\fref{fig:v-level_inversion}(a)], we see that for a given $\Delta_2$ the minimum is always at $\Delta_3 = \Delta_2$. The scan over different Rabi frequencies [\fref{fig:v-level_inversion}(b)] shows that there is a threshold which the driving amplitudes have to surpass in order to achieve population inversion. Yet, there is also an upper limit: if either Rabi frequency becomes much larger than all other frequencies in the system, one simply obtains the result of a strongly driven two-level transition; i.e., the population is distributed equally between the ground state and the strongly driven upper level. If both Rabi frequencies become extremely large simultaneously, half the population accumulates in the ground state, while the excited states are populated with a quarter each.

Note also, that the symmetry in the detunings is just due to the freedom of choice zero-point energy (direction of the rotating frame; $\tilde{\Delta}_i \rightarrow -\Delta_i$).

\subsection{Effects of Driving with a Finite Laser Linewidth}

\begin{figure}
\centering
\begin{minipage}[b]{0.48\linewidth}
\centering
\includegraphics[width=\linewidth]{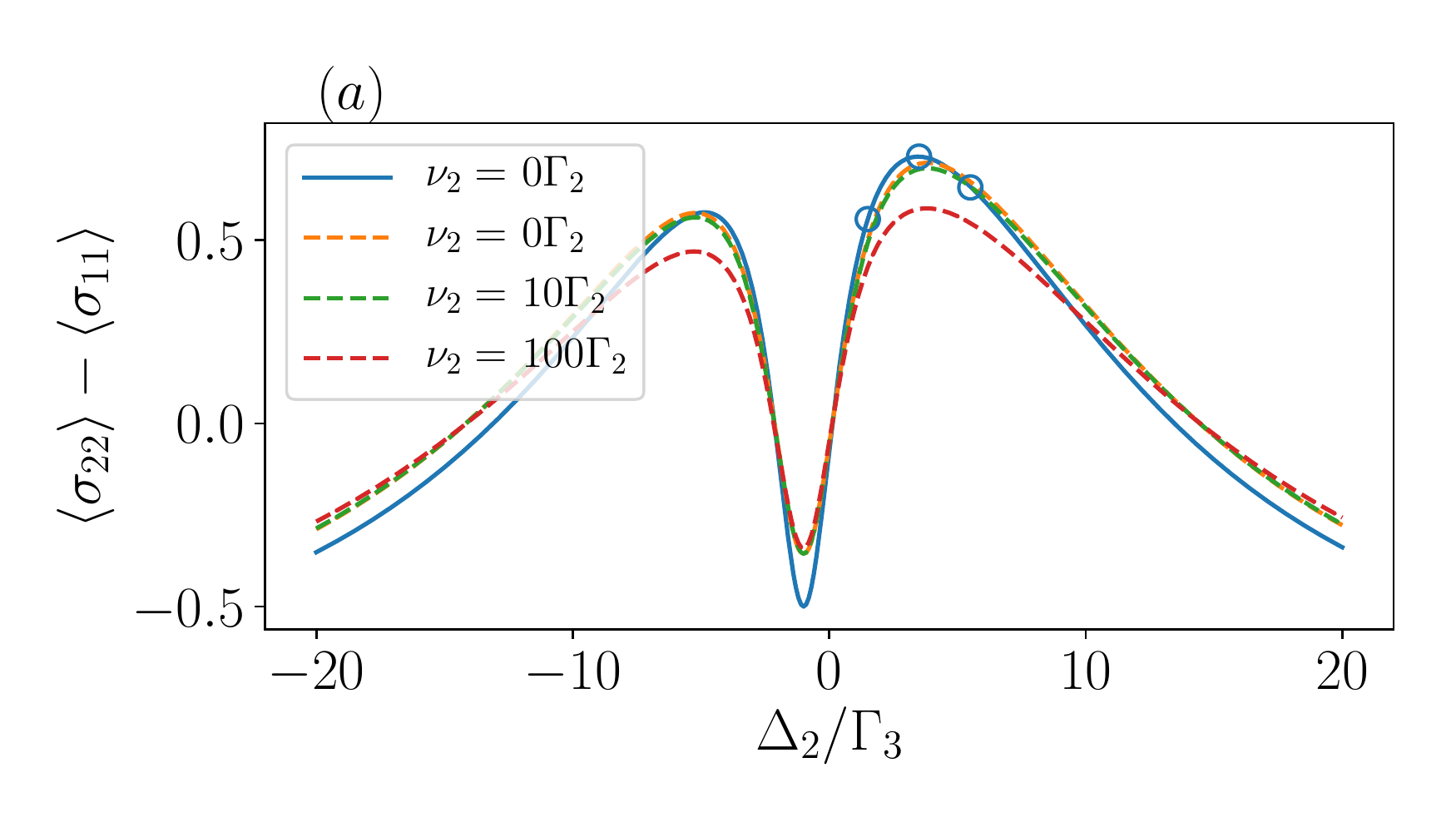}
\end{minipage}
\begin{minipage}[b]{0.48\linewidth}
\centering
\includegraphics[width=\linewidth]{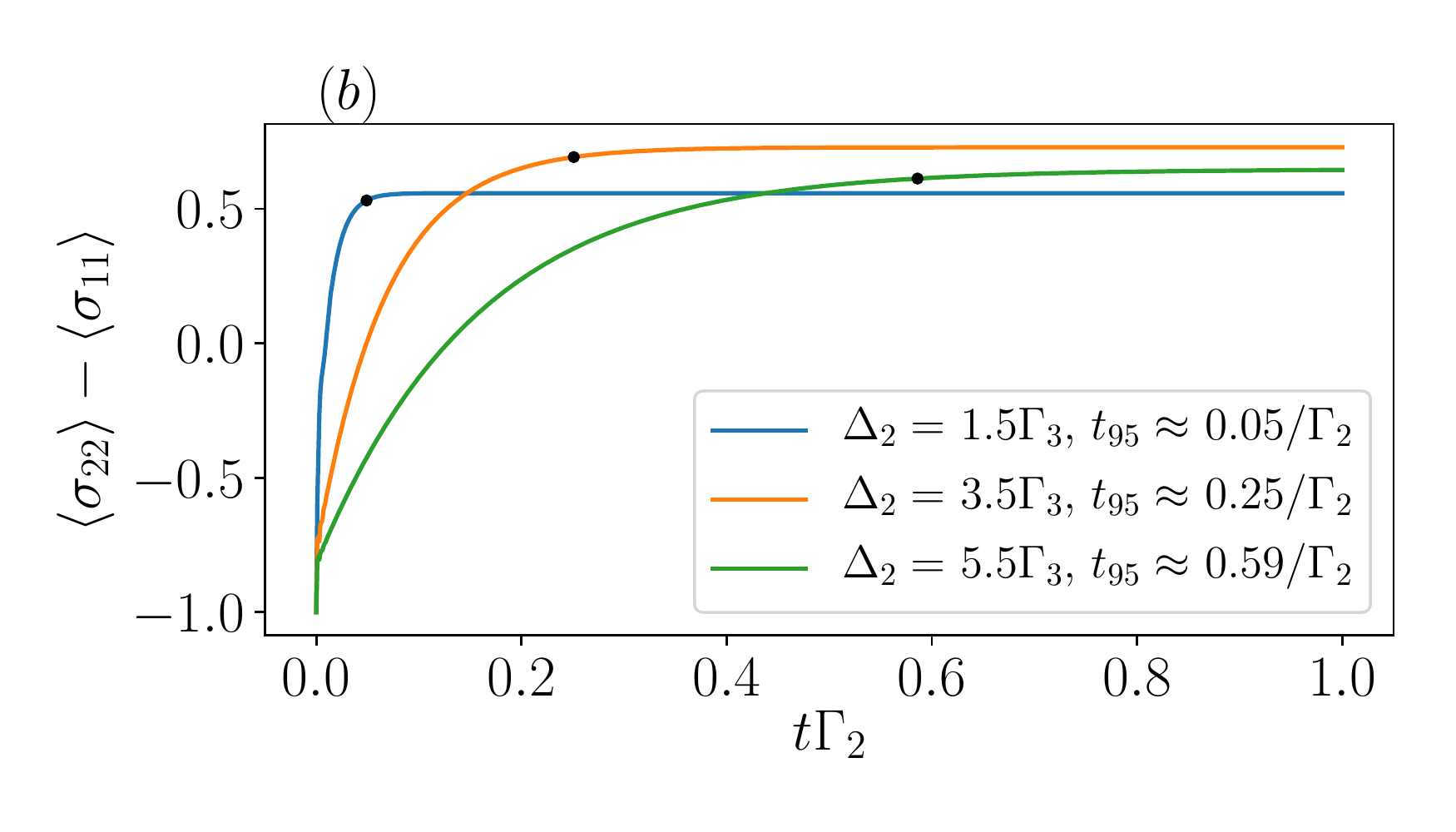}
\end{minipage}
\caption{\emph{Laser linewidth influence and process time scale}. In figure (a) we can see that for $\nu_2 = 10\Gamma_2$ the population difference $\braket{\sigma_{22}} - \braket{\sigma_{22}}$ is still almost the same. The blue solid line is for $\nu_3 = 0$ and the dashed lines are for $\nu_3 = 0.5 \Gamma_3$. Since $\Gamma_3 \gg \Gamma_2$, the linewidth $\nu_3$ can be neglected in comparison to $\nu_2$. The parameters are $\Delta_3 = -\Gamma_3$, $\Omega_2 = 0.5 \Gamma_3$, $\Omega_3 = 0.5 \Gamma_3$ and the blue circles indicate the values for $\Delta_2$ in (b). Figure (b) shows that the steady state can be reached much faster if $\Delta_2$ is closer to zero. Here, we have $\nu_2 = \nu_3 = 0$, and the black dots indicate the $t_{95}$ data points.}
\label{fig:laser_linewidth_timescale}
\end{figure}

The driving term in the Hamiltonian \eqref{eq:Hamiltonian} assumes lasers with an infinitely small linewidth. Certainly, this is not always a good assumption for real experimental setups. In particular, the linewidth of transitions in clock atoms (such as the ones considered here) can be much smaller than that of a driving laser. Thus, we study the influence of a finite pump laser linewidth on the system. A finite linewidth $\nu_i$ of the driving laser on the transition $| 1 \rangle \leftrightarrow | i \rangle$ can be modelled by an effective atomic decoherence process in the form of dephasing \cite{Plankensteiner2016laser_noise, Dorner2012quantum_frequency}. The Liouvillian for such a process is
\begin{align}\label{eq:dephasing}
\mathcal{L}_{\nu_i}[\rho] = \frac{\nu_i}{2} (2 \sigma_{ii} \rho \sigma_{ii} - \sigma_{ii} \rho - \rho \sigma_{ii}).
\end{align}
In \fref{fig:laser_linewidth_timescale}(a), we see that for laser linewidths up to the order of $10\Gamma_2$ the population difference stays almost the same. This is a consequence of the Rabi frequency $\Omega_2$ being much larger than the pump laser linewidth $\nu_2$. Only with a dephasing at the order of $100\Gamma_2$, we start to see a significant reduction of the maximal population inversion. Hence, the laser on the narrower transition does not need to be extremely narrow in order to excite the atom, which is advantageous in experimental setups. Furthermore, a linewidth of the laser on the broader transition of up to several hundred $\Gamma_2$ ($\nu_3 = 0.5\Gamma_3 \approx 2000\Gamma2$) has almost no impact on the state population. This is expected since $\Gamma_3 \gg \Gamma_2$.

\subsection{Time scale}

For some applications it is necessary or useful to create the population inversion on a shorter timescale. In \fref{fig:laser_linewidth_timescale}(b) the time evolution of the population difference for different values of $\Delta_2$ is shown. First of all we can see that the timescale is determined by $\Gamma_2$ rather than $\Gamma_3$. Thus, the time needed to reach the steady-state population inversion can be quite long, e.g. for clock atoms with decay rates in the mHz-regime. Furthermore, we also see that the time strongly depends on $\Delta_2$. The closer $\Delta_2$ is to zero, the faster the steady state can be reached. However, this can cause a lower population inversion.

In order to quantitatively compare the times needed to reach the steady state, we introduce the variable $t_{95}$, which is the time at which the population inversion exceeds $95 \%$ of its steady-state value. By choosing the lowest possible value for $\Delta_2$ we can reduce $t_{95}$ by up to one magnitude compared to the case with the highest inversion. To further decrease $t_{95}$ we can establish a bigger effective decay rate $\Gamma_2^{\text{eff}} > \Gamma_2$, which then determines the time scale. This can be achieved by, for example, creating an additional decay channel from $| 2 \rangle$ to $| 1 \rangle$. Since $| 3 \rangle$ decays into $| 1 \rangle$ with $\Gamma_3 \gg \Gamma_2$ we could open an additional decay channel for $\ket{2}$ if it is possible to incoherently drive the atom from $| 2 \rangle$ to $| 3 \rangle$. The corresponding rate needs to be $\Gamma_{23} \approx (M -1 )/M\Gamma_2$ in order to have $M$-fold faster decay from $\ket{2}$. This is valid as long as $\Gamma_2^{\text{eff}} \ll \Gamma_3$ holds.


\section{Continuous Stationary Lasing} \label{sec:laser}

We proceed by considering an ensemble of V-type atoms placed inside an optical resonator, as shown in \fref{fig:schematic}(b). Since inversion on the narrow transition can be achieved using the driving scheme, the atoms act as gain for the field inside the resonator. In the following, we investigate the properties of the output light and show that we obtain continuous lasing.

Consider $N$ V-level atoms inside a cavity, each of which couples with a rate $g_j$ to the cavity field via the transition $\ket{1}\leftrightarrow\ket{2}$. The Hamiltonian is given by
\begin{equation}
H = -\Delta_{\mathrm{c}} a^{\dagger}a + \sum_{j=1}^N g_j (a^{\dagger} \sigma_{12}^{j} + a \sigma_{21}^{j}) + \sum_{ i=\{2,3\}} \sum_{j=1}^N -\Delta_i^{j} \sigma_{ii}^{j} + \Omega_i^{j} (\sigma_{i1}^{j} + \sigma_{1i}^{j}),
\label{eq:Hamiltonian_lasing}
\end{equation}
where $\Delta_{\mathrm{c}} = \omega_{\ell 2} - \omega_{\mathrm{c}}$ is the detuning between the cavity resonance frequency $\omega_{\mathrm{c}}$ and the laser frequency $\omega_{\ell 2}$. The cavity photon creation (annihilation) operator is denoted by $a^{\dagger}$ ($a$) and the superscript index $j$ specifies the $j$-th atom. Photons leaking through the cavity mirrors at a rate $2 \kappa$ give rise to an additional Liouvillian term
\begin{equation}
\mathcal{L}_\kappa[\rho] = \kappa (2 a \rho a^\dagger - a^\dagger a \rho - \rho a^\dagger a).
\label{eq:cavity_decay}
\end{equation}
For individually decaying atoms, the same decay process as described in \eqref{eq:liouvillian_decay} applies to each atom. Thus, we have
\begin{equation}
\mathcal{L}_\mathrm{N\Gamma}[\rho] = \sum_{i = \{ 2,3 \}} \frac{\Gamma_i }{2} \sum_{j=1}^N (2 \sigma_{1i}^j \rho \sigma_{i1}^j - \sigma_{ii}^j \rho - \rho \sigma_{ii}^j),
\label{eq:ind_decay_sum}
\end{equation}
for the decay processes of both excited states. If we assume that all atoms are driven by the same laser, we obtain the following dissipative processes due to the finite laser linewidth (see \ref{sec:dephasing_pump} for details). On the transition $| 1 \rangle \leftrightarrow | 3 \rangle$ we get a dephasing with 
\begin{equation}
\mathcal{L}_{N\nu_3}[\rho] = \frac{\nu_3}{2} (2 S_3 \rho S_3 - S_3^2 \rho - \rho S_3^2),
\label{eq:dephasing3_N}
\end{equation}
whereas on the transition $| 1 \rangle \leftrightarrow | 2 \rangle$ we have
\begin{equation}\label{eq:dephasing2_N}
\mathcal{L}_{N\nu_2}[\rho] = \frac{\nu_2}{2} (2 (a^\dagger a + S_2) \rho (a^\dagger a + S_2) - (a^\dagger a + S_2)^2 \rho - \rho (a^\dagger a + S_2)^2),
\end{equation}
where $S_{i} = \sum_{j=1}^N \sigma_{ii}^j$ is the collective atomic operator.
The additional term ($a^\dagger a$) in \eqref{eq:dephasing2_N} is due to the shared rotating frame of the atom and the cavity. The full Liouvillian for the lasing setup then reads
\begin{equation}
\mathcal{L}_\mathrm{laser}[\rho] = \mathcal{L}_\kappa[\rho] + \mathcal{L}_{N\Gamma}[\rho] + \mathcal{L}_{N\nu_2}[\rho] + \mathcal{L}_{N\nu_3}[\rho]
\label{eq:liouvillian_lasing}
\end{equation}

Solving the master equation for more than just a few atoms is an impossible task due to the exponential scaling of the Hilbert space with the atom number. Therefore, we employ a second-order cumulant expansion \cite{Kubo1962generalized} to calculate the time evolution of average values of interest. Furthermore, we assume that all atoms couple equally to the cavity. Exploiting the symmetry of the system renders $N$ a constant factor, which does not change the number of equations one needs to solve. This allows us to solve the equations of motion for a large number of atoms. The most relevant second-order equations can be found in appendix \ref{sec:sys_eqs}.

\subsection{Lasing Threshold}

As a first step, we compute the normalized power spectral density $S(\omega)$ of a single atom when it is subject to the driving scheme. This gives us a general idea of what to expect for the resulting lasing output.

\begin{figure}[ht]
\centering
\centering\includegraphics[width=0.6\linewidth]{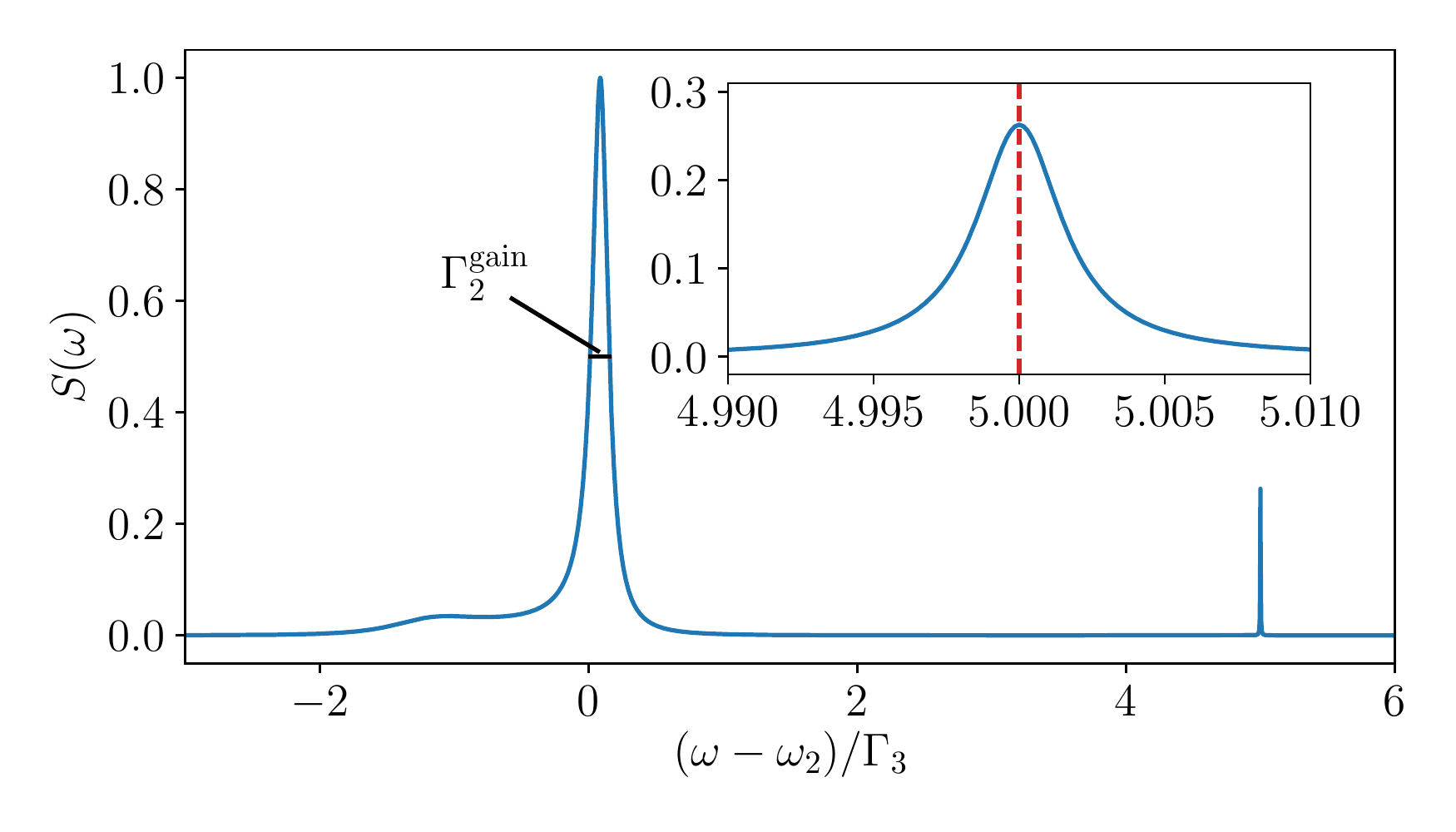}
\caption{\emph{Single atom emission spectrum}. The peak at the atomic resonance frequency ($\omega = \omega_2$) has a FWHM of approximately $\Gamma_2^\text{gain} = 614\Gamma_2$, and the smaller peak at the pump laser frequency ($\omega = \omega_{\ell 2} = 5\Gamma_3 + \omega_2$, see inset) of $\text{FWHM} \approx 15\Gamma_2$. The red, dashed line in the inset indicates the driving laser frequency $\omega_{\ell 2}$. The parameters are $\Delta_2 = 5\Gamma_3$, $\Delta_3 = -1 \Gamma_3$, $\Omega_2 = 0.5 \Gamma_3$, $\Omega_3 = 0.5 \Gamma_3$ and $\nu_2 = \nu_3 = 0$. This single atom emission spectrum was calculated with a master equation approach, using the Wiener-Khinchin theorem \cite{puri2001mathematical}.}
\label{fig:atom_spectrum}
\end{figure}

Due to the strong coherent driving amplitude, a considerable power broadening on the otherwise narrow transition is induced. This transforms the naturally narrow gain medium into a relatively broad one, see \fref{fig:atom_spectrum}. Thus, with respect to the power-broadened linewidth, the resulting laser operates in the good-cavity regime ($\kappa \ll \Gamma_2^\text{gain}$). In general, the strong coherent drives lead to distinct energies of multiple dressed states, i.e. they induce considerable ac-Stark shifts. For example, for the parameters chosen in \fref{fig:atom_spectrum}, we see that the laser gain peak is slightly shifted from the bare resonance frequency $\omega_2$. Furthermore, we observe an additional small and broad peak to the left of the laser gain frequency, which is the signature of a dressed state. The other small, but narrow peak (linewidth $\sim 15\Gamma_2$) in \fref{fig:atom_spectrum} (see inset) is located at the frequency of the pump laser ($\omega=\omega_{\ell 2}$) and can therefore not be used for lasing. A large amount of photons from the pump laser would be coherently scattered into the cavity. At the same time, the largest amount of emitted power is far detuned from any driving laser. Coherent scattering of the driving laser into the cavity is therefore suppressed. This already indicates that lasing can indeed be achieved at this frequency.

\begin{figure*}[t]
\begin{minipage}[b]{0.45\linewidth}
\centering
\includegraphics[width=\textwidth]{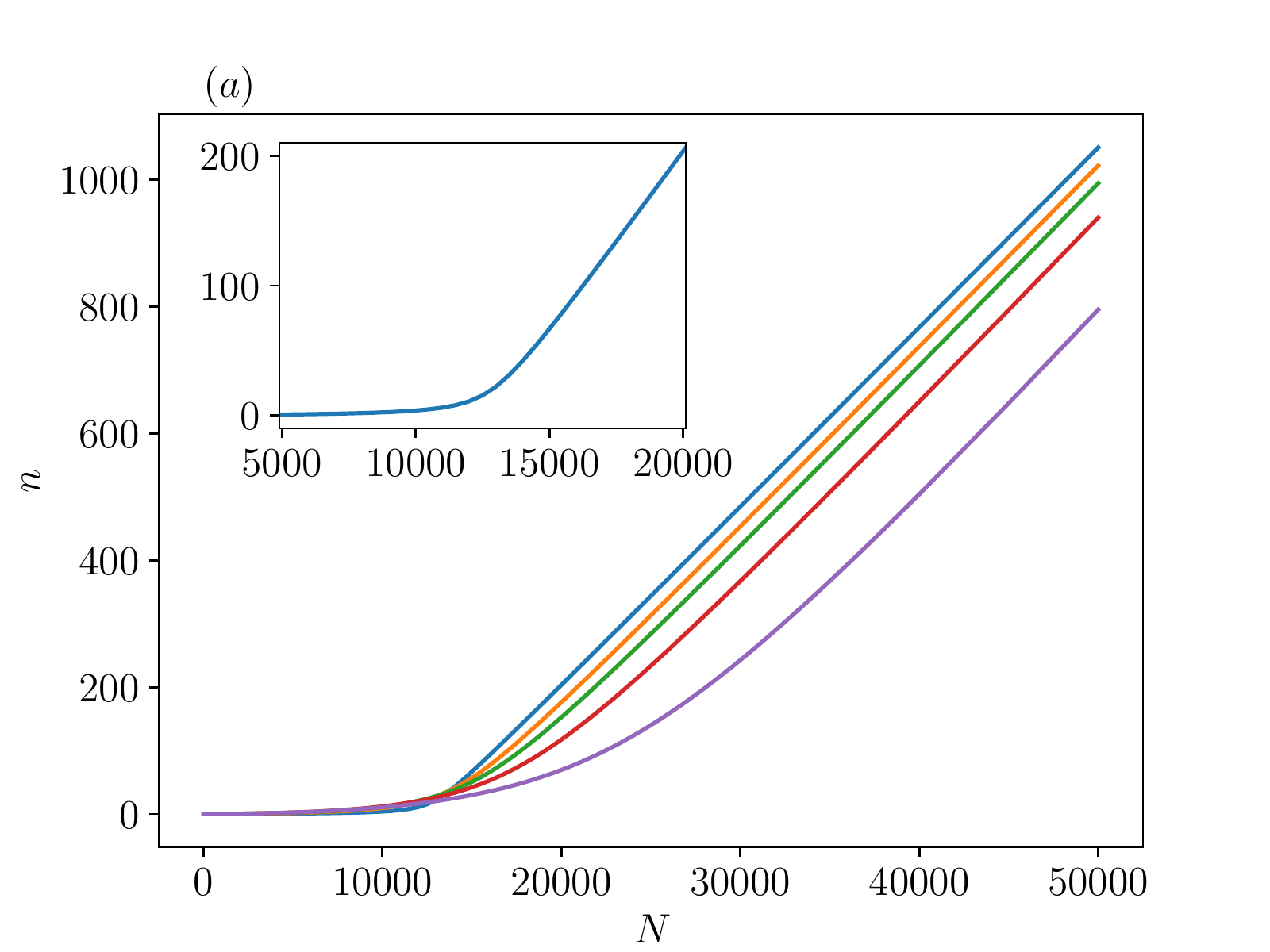}
\end{minipage}
\hspace{0.2cm}
\begin{minipage}[b]{0.45\linewidth}
\centering
\includegraphics[width=\textwidth]{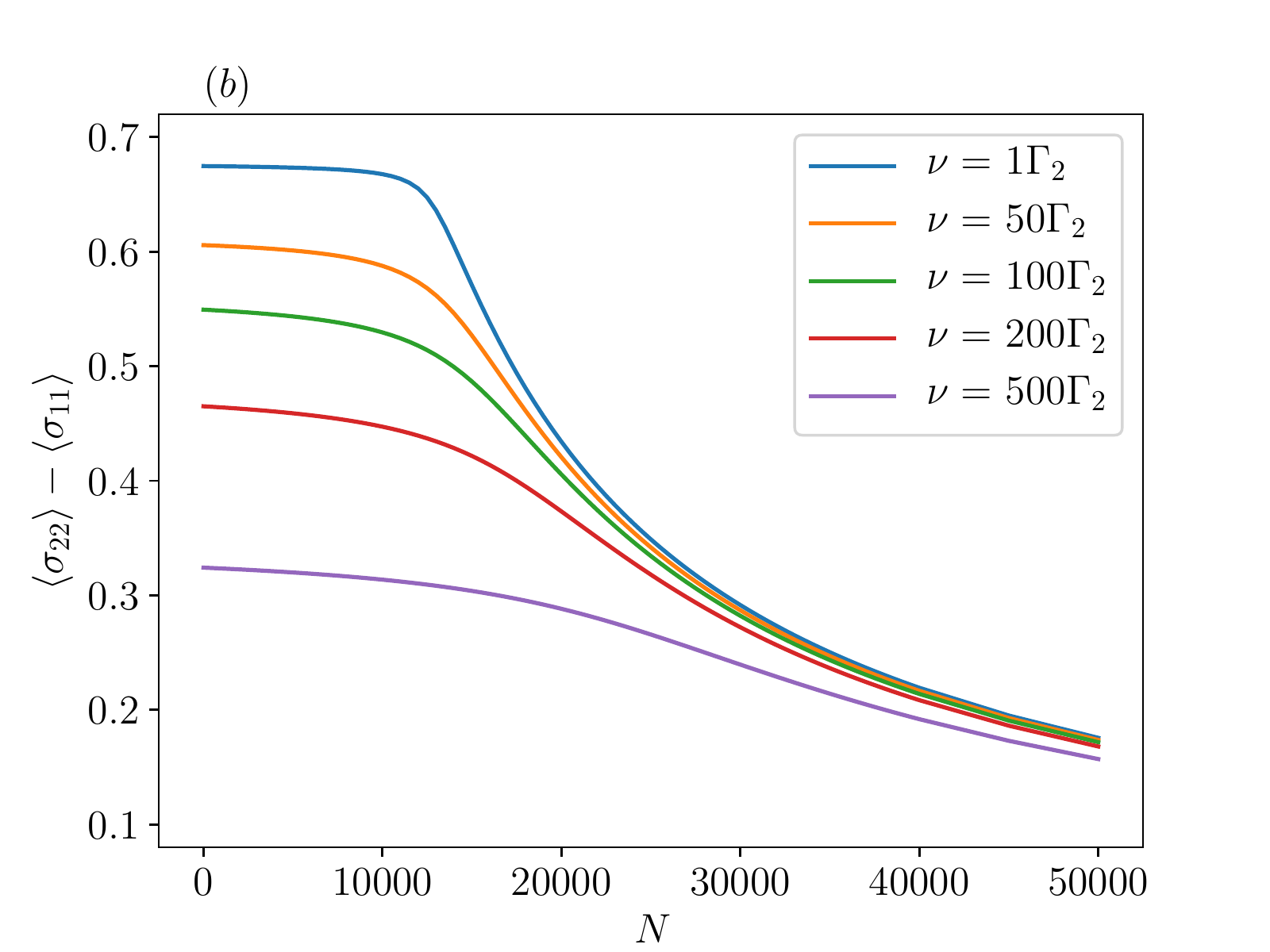}
\end{minipage}
\caption{\emph{Laser threshold behaviour}. The threshold behaviour of the cavity photon number $n$ on the atom number $N$ is shown in the figure (a) and the population inversion $\langle \sigma_{22} \rangle - \langle \sigma_{11} \rangle$ for the same dephasings ($\nu \equiv \nu_1 = \nu_2$) and number of atoms is shown in (b). The parameters are $\Delta_2 = 5\Gamma	_3$, $\Delta_3 = -1 \Gamma_3$, $\Omega_2 = 0.5 \Gamma_3$, $\Omega_3 = 0.5 \Gamma_3$, $\Delta_{\mathrm{c}} = \Delta_2$ ($\omega_{\mathrm{c}} = \omega_2$), $g = 2\Gamma_2$ and $\kappa = 50\Gamma_2$.}
\label{fig:threshold}
\end{figure*}

Making use of the inversion scheme on the narrow transition results in a steady state of the optical cavity featuring a potentially large number of photons. Therefore, the optical resonator provides a continuous output. In \fref{fig:threshold} we investigate the behaviour of the system with an increasing number of atoms $N$. We find that the system exhibits a threshold, as can be clearly seen in \fref{fig:threshold}(a), where we plot the steady-state photon number inside the cavity. Once the threshold is passed, the number of photons inside the cavity (and hence the lasing power) rapidly increases. The threshold atom number is approximately $N \approx 12000$ for $\nu_1=\nu_2 = \Gamma_2$ (see inset). For stronger dephasing a larger gain medium is required in order to sustain the lasing operation. Therefore, the number of atoms needed to pass the threshold increases. This can also be seen from \fref{fig:threshold}(b), where the population inversion per atom is depicted. For stronger dephasing, the inversion decreases and thus the gain provided by each atom is reduced. The key observation, however, is that the threshold can still be passed almost regardless of the linewidth of the driving lasers used in the inversion scheme.

\begin{figure*}[t]
\begin{minipage}[b]{0.45\linewidth}
\centering
\includegraphics[width=\textwidth]{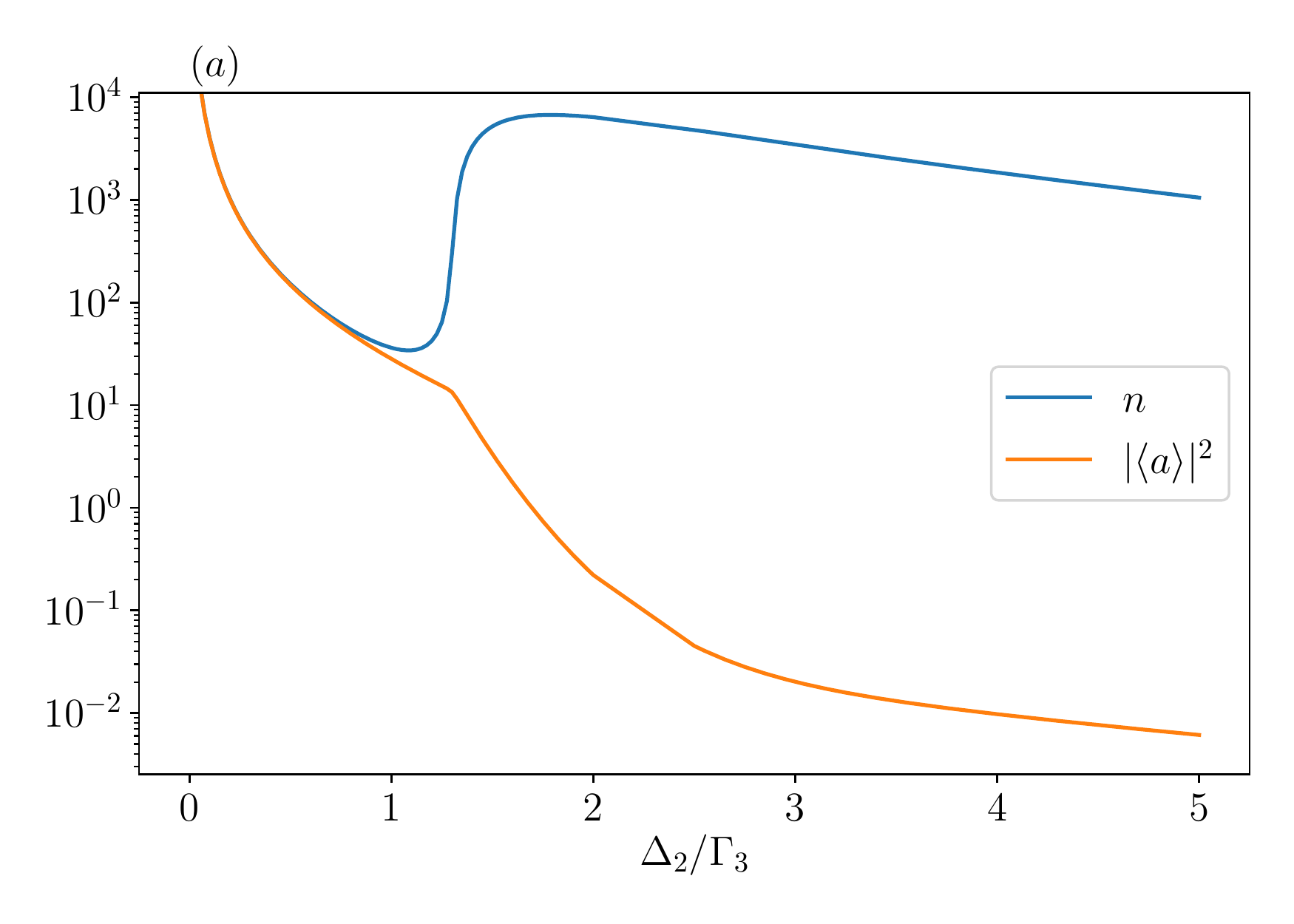}
\end{minipage}
\hspace{0.2cm}
\begin{minipage}[b]{0.45\linewidth}
\centering
\includegraphics[width=\textwidth]{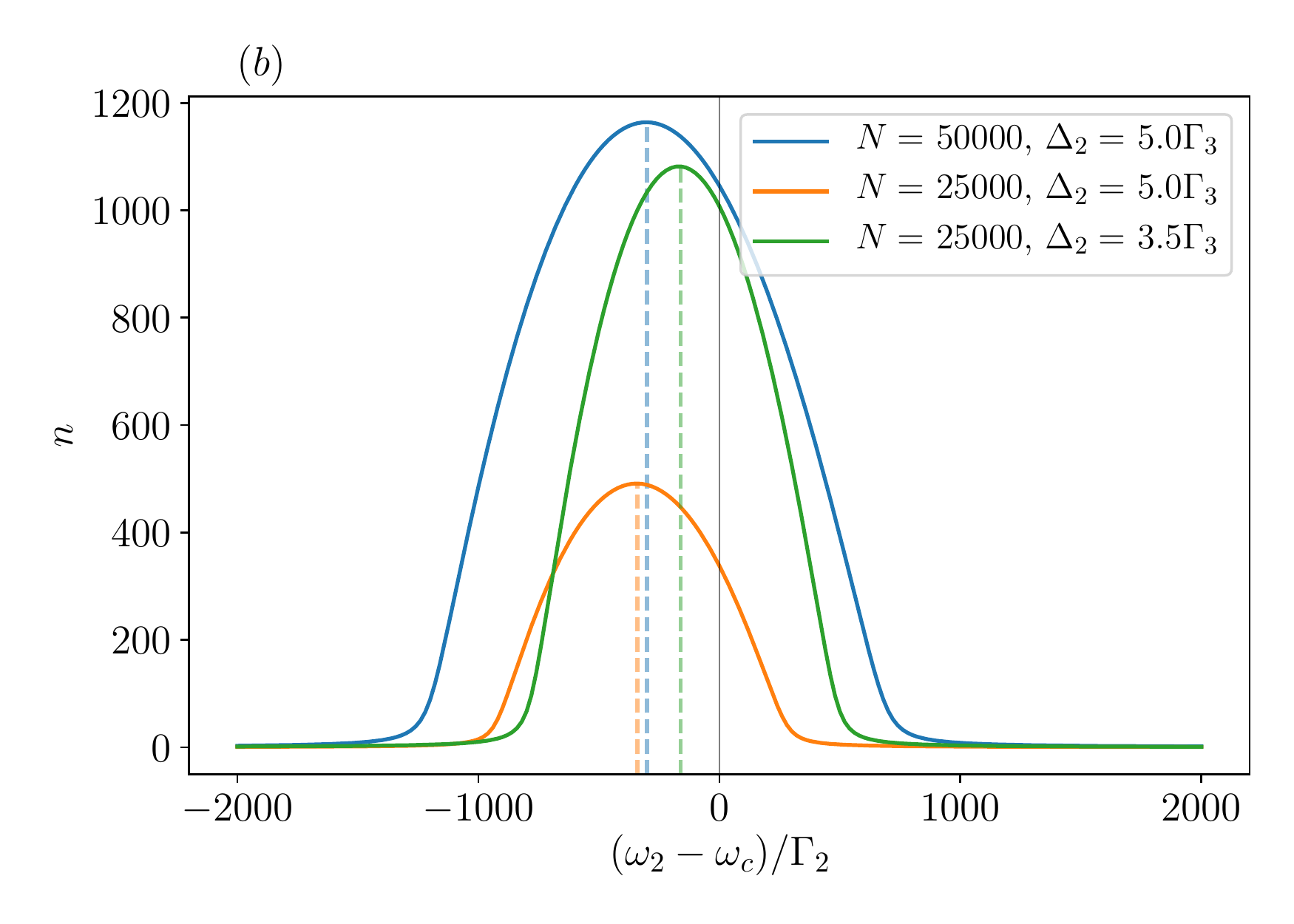}
\end{minipage}
\caption{\emph{Stimulated emission of photons}. Figure (a) shows the steady-state average photon number $n$ and the coherent fraction of photons $| \langle a \rangle |^2$ inside the cavity as a function of the detuning $\Delta_2$. The cavity is always on resonance with the unperturbed atomic transition frequency ($\Delta_{\mathrm{c}} = \Delta_2$). If $\Delta_2$ is small, coherent scattering of photons into the cavity is more likely to occur. In figure (b) the photon number $n$ is plotted as a function of the detuning between the atomic transition frequency and the cavity resonance frequency $\omega_2 - \omega_{\mathrm{c}} = \Delta_{\mathrm{c}} - \Delta_2$. For the maximum photon number the cavity needs to be blue detuned from the atoms ($\omega_{\mathrm{c}} > \omega_2$). The parameters when kept constant for both subfigures are the same as in \fref{fig:threshold} for $N = 50000$.}
\label{fig:photons_Delta2_Deltac}
\end{figure*}

In order to avoid coherent scattering of photons from the pump laser on the transition $| 1 \rangle \leftrightarrow | 2 \rangle$ into the cavity, we need to ensure that the laser is far detuned from the cavity resonance frequency. Because of this, we chose a large detuning of $\Delta_2 = 5\Gamma_3$, at which the inversion scheme also works well. \fref{fig:photons_Delta2_Deltac}(a) shows the amount of coherently scattered photons $| \langle a \rangle |^2$ in comparison to the total photon number $n$ in the cavity as a function of $\Delta_2$. We see, that if the detuning $\Delta_2$ is small, a considerable amount of photons enter the cavity via coherent scattering. Moreover, we can see that the chosen value for $\Delta_2$ is not optimal. The cavity photon number therefore is not maximal. If we change $\Delta_2$ from $5\Gamma_3$ to e.g. $2\Gamma_3$, the photon number would in fact increase by almost one order of magnitude, still keeping the coherently scattered photon number sufficiently low.

\fref{fig:photons_Delta2_Deltac}(b) shows that, if the cavity is blue detuned from the atomic transition frequency, the photon number can further increase. Of course, if the cavity is too far off-resonant, the photon number almost vanishes. The three different sets of parameters considered, indicate that above threshold an increasing atom number $N$ and a decreasing pump laser detuning $\Delta_2$ shift the optimal cavity resonance frequency towards the atomic resonance frequency.

\subsection{Cavity Emission Spectrum}  \label{sec:laser_spec}

The steady-state cavity power spectral density can be calculated as the Fourier transform of the first order correlation function $g^1(\tau) = \langle a^\dagger(\tau) a(0)$ 
\begin{equation}
S(\omega) = \int_0^\infty d\tau g^1(\tau) e^{-i(\omega)\tau} .
\label{eq:wiener-khinchin}
\end{equation}
Using the quantum regression theorem \cite{carmichael2013statistical} we can calculate the time evolution of the correlation function $g^1(\tau)$ with a second order cumulant expansion. In appendix \ref{sec:dephasing_pump} one can see that the time evolution for the laser systems with finite pump laser linewidth have been calculated in a fluctuating rotating frame ($U(t) = e^{i(\omega_{\ell 2}t + \phi(t))(a^\dagger a + \sum_{j=1}^N \sigma_{22}^j)}$, instantaneous frame). But since we want to obtain the spectrum with respect to a stable monochromatic reference frequency we are only allowed to transform the system into a non-fluctuating rotating frame (e.g. $U(t) = e^{i \omega_{\ell 2}t(a^\dagger a + \sum_{j=1}^N \sigma_{22}^j)}$, coherent frame) \cite{avan1977two}. The set of equations to calculate the correlation function and a detailed derivation is shown in appendix  \ref{sec:corr_fct_eqs} and \ref{sec:corr_deriv}, respectively. 

\begin{figure*}
\begin{minipage}[b]{0.28\linewidth}
\centering
\includegraphics[width=\textwidth]{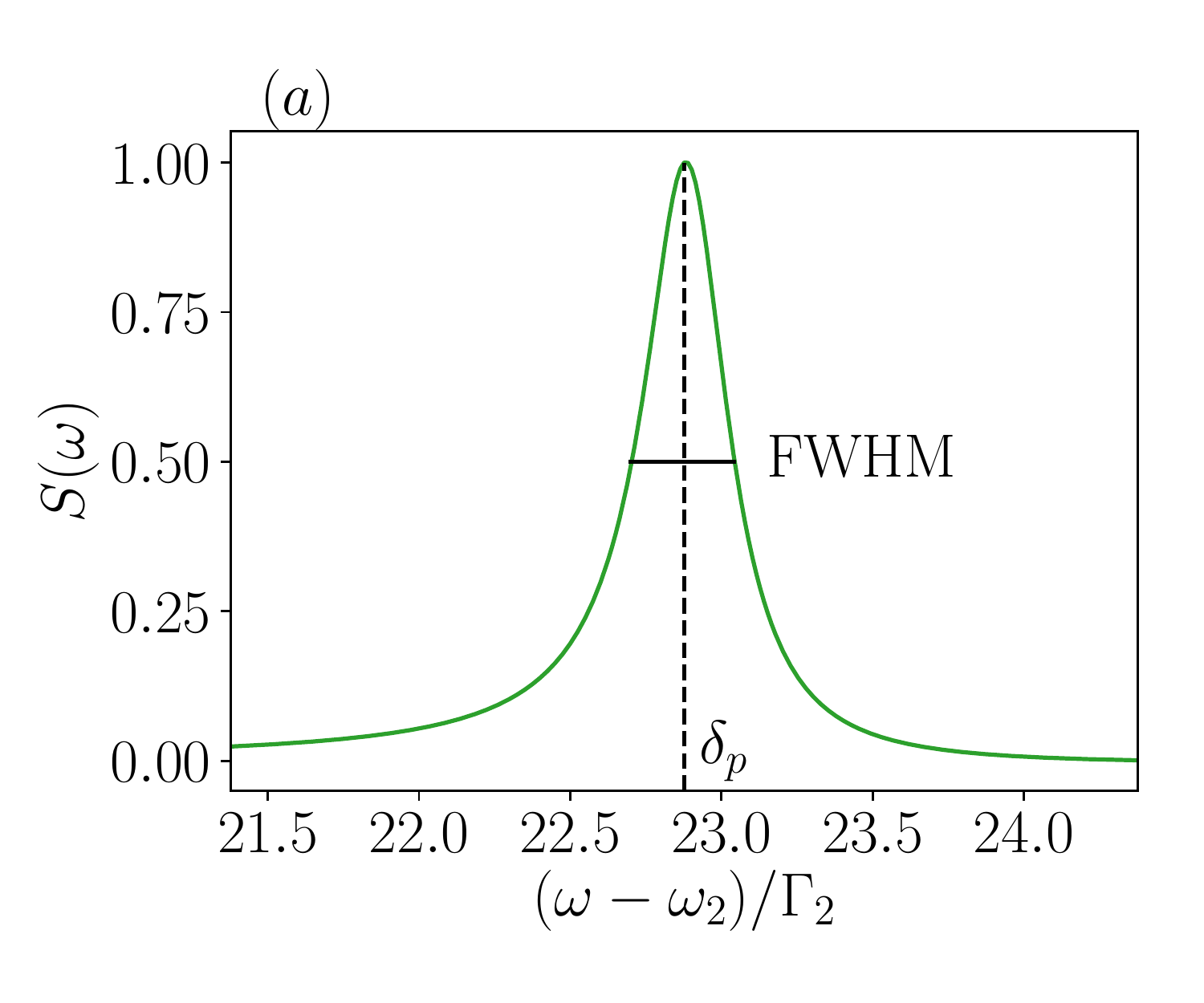}
\end{minipage}
\begin{minipage}[b]{0.32\linewidth}
\centering
\includegraphics[width=\textwidth]{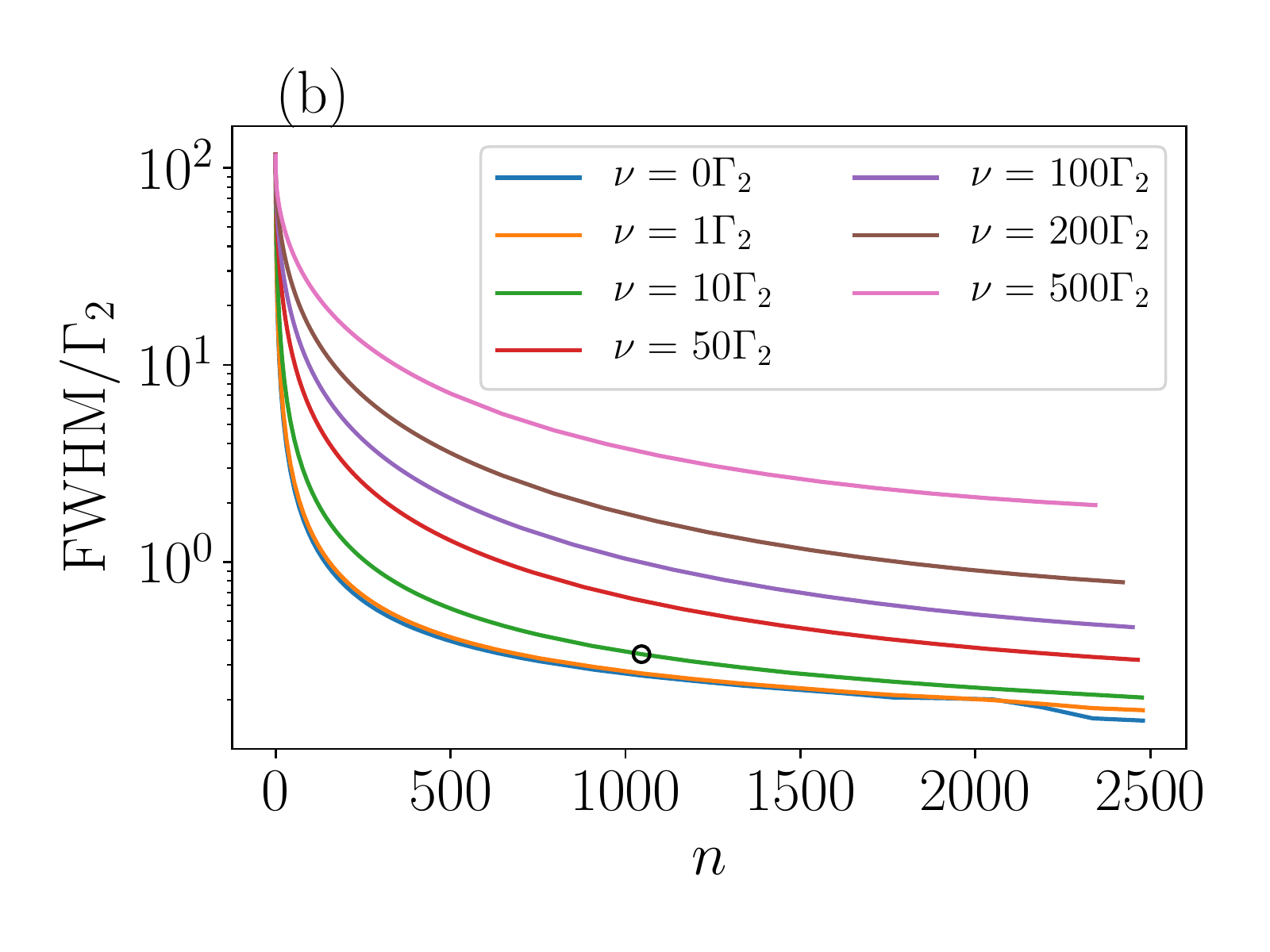}
\end{minipage}
\begin{minipage}[b]{0.32\linewidth}
\centering
\includegraphics[width=\textwidth]{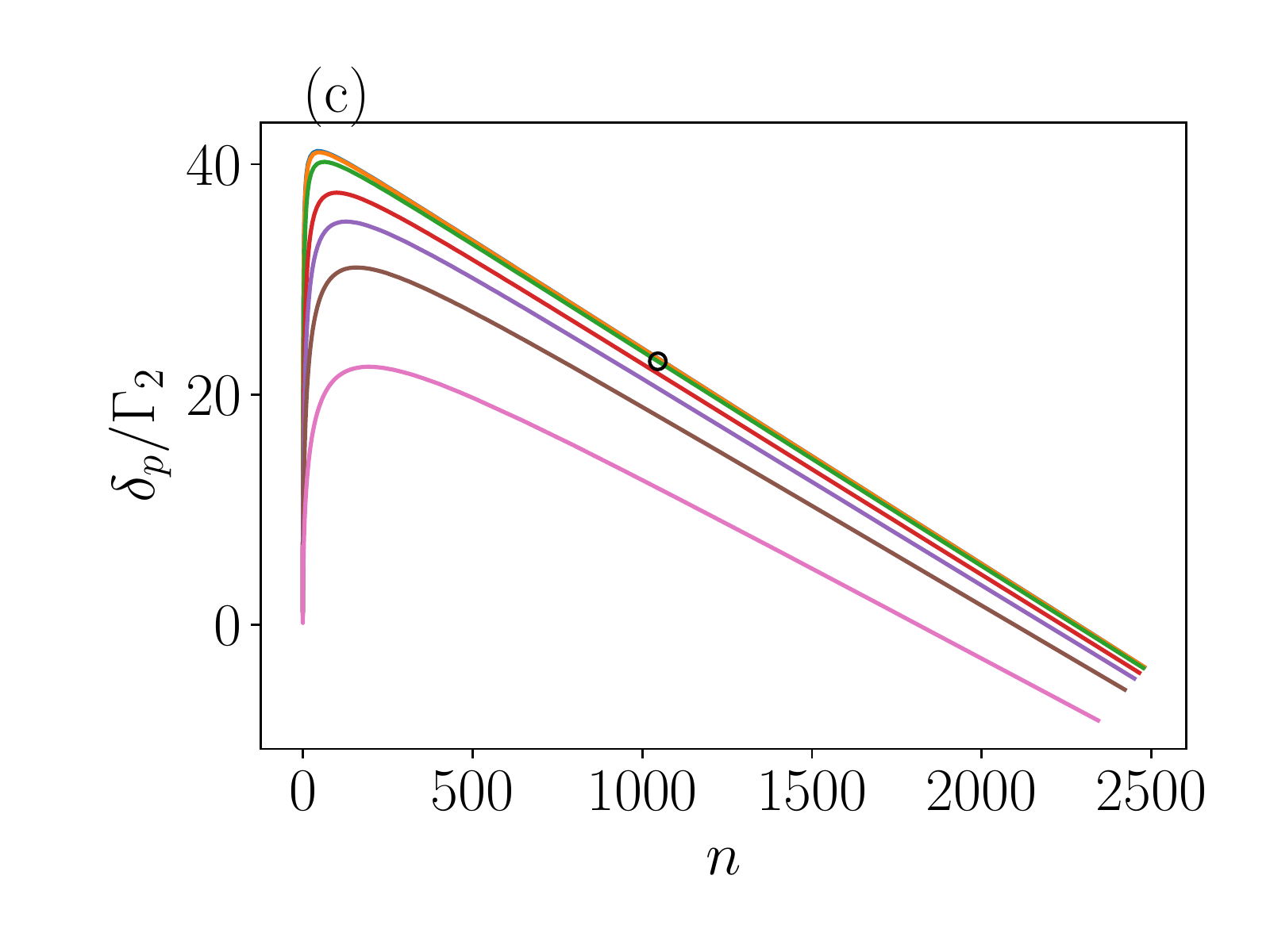}
\end{minipage}
\caption{\emph{Spectral properties of the cavity}. Figure (a) shows a typical (normalized) spectrum $S(\omega)$ above threshold, where we chose $N=50000$ and $\nu = 10\Gamma_2$ (black circles in (b) and (c)). The FWHM ($0.35\Gamma_2$) and the peak position $\delta_p$ $(22.9 \Gamma_2)$ are indicated. Note that the spectrum is plotted with respect to the atomic transition frequency. The reduction of the laser linewidth with increasing cavity photon number is plotted in (b). And (c) shows the shift of the peak position due to a stark shift caused by large cavity photon numbers. The legend in (b) is for all three plots.}
\label{fig:fwhm_peak_pos}
\end{figure*}

The properties of the cavity emission spectrum are depicted in \fref{fig:fwhm_peak_pos}. The most significant result here is that a FWHM below $\Gamma_2$ can be reached, even if the linewidths of the driving lasers are above $100\Gamma_2$. Hence, a narrow bandwidth laser can be achieved with relatively broad pump lasers. Furthermore, the system does not rely on a direct decay channel into the lasing transition, as opposed to conventional laser systems. Only a V-level structure is necessary, which can be often found in rare earth atoms, commonly used in optical clocks. Let us stress here, that we assumed an ideal model with all atoms fixed at the cavity field anti-nodes.

As expected for a conventional laser, the FWHM is approximately given by the cavity linewidth $2\kappa$ for small photon numbers [see \fref{fig:fwhm_peak_pos}(b)]. Above threshold, where the photon number is large, it reduces with $\text{FWHM}(n) \sim 1/n$. This behaviour is well-know in the case of good-cavity lasers \cite{schawlow1958infrared}. In \fref{fig:fwhm_peak_pos}(c), we can see that the peak position shows an almost linear dependency on the photon number, if the latter is large. This is caused by an ac-Stark shift due to the cavity field. In general we find that the spectral properties are very similar to those of a conventional laser. For example, also the cavity pulling coefficient above threshold is $d\delta_p(\Delta_{\mathrm{c}}) / d\Delta_{\mathrm{c}} \approx 1$.

The FWHM shows an unexpected behaviour for $\nu = 0$ at high photon numbers. The most likely reason for this is that the second order cumulant expansion reaches its limits there. The fact that this happens only for $\nu = 0$ is a good indicator: the dephasing destroys the coherences and therefore makes the system more classical, i.e. the approximation is more accurate. By keeping specific third order terms as e.g. $\langle a^\dagger a \sigma_{22} \rangle$ \cite{henschel2010cavity} one may get rid of this inaccuracy. However, the equations listed in the appendix are already quite lengthy; taking third order corrections would ultimately go beyond the intended scope of this work.

\section{Cooling}

Another point of interest for the stability of continuous lasing operation is cooling of the atoms. On the one hand, this ensures that the atoms stay in the cavity as long as possible thereby providing gain. On the other hand, heating will generally cause noise and broadening of the gain medium, which will have a detrimental impact on the spectral properties of the laser.

Therefore, we consider the impact of the inversion scheme on the atomic motion in this section. To this end, effects of photon recoil from spontaneous emission events are taken into account. This allows us to estimate the temperature of the gain medium when subjected to the two driving lasers.

We employ a Monte-Carlo wave function (MCWF) approach \cite{Dum1992montecarlo, Molmer1993MonteCarlo}, where we treat the atomic motion as classical variables. In the MCWF method, the norm of the state vector decreases over time. Once the norm decreases below a certain (randomly chosen) value, a quantum jump occurs. In our approach, we include an additional momentum kick whenever a jump occurs.

In the following, we restrict the atomic motion to two spatial dimensions. The pump lasers are considered to be aligned perpendicularly, in directions determined by their respective wavevectors $\textbf{k}_i$, respectively (i.e. $\textbf{k}_2\cdot\textbf{k}_3=0$). Furthermore, we neglect the influence of the cavity field in these calculations. In the case of many photons inside the cavity this assumption may be far from ideal. However, note that the lasing operation is optimal (maximal output power) when the cavity is blue detuned from the atomic transition frequency, as shown in [\fref{fig:photons_Delta2_Deltac}(b)]. Therefore, the cavity would effectively provide an additional cooling mechanism \cite{salzburger2004atomic, hotter2019superradiant}. Thus, neglecting the cavity leads to higher final temperatures, which provides a sufficient estimate.

Since the atoms do not interact with one another, we repeatedly compute trajectories of the particle motion for a single atom only. The considered system is modeled as follows: the internal atomic structure is treated quantum mechanically, while the motional degrees of freedom are assumed to be fully described by their average values. This assumption is well justified if the momentum of the atom is large compared to the momentum of a single photon. The Hamiltonian of this system reads
\begin{equation}
H = \sum_{ i=\{2,3\}} -\Delta_i \sigma_{ii} + \Omega_i \cos(\textbf{k}_i \cdot \textbf{r}) (\sigma_{i1} + \sigma_{1i}),
\label{eq:Hamiltonian_sc}
\end{equation}
where $\textbf{r}$ is the atomic position vector in the plane spanned by the wavevectors. The atomic motion is described by the classical equations of motion for the velocity
\begin{align}
\dot{\textbf{r}} &= \frac{\textbf{p}}{m},
\end{align}
and the forces acting on the atom
\begin{align}
\dot{p}_i &=  k_i \Omega_i \sin(\textbf{k}_i \cdot\textbf{r}) 2 \Re \braket{\sigma_{1i}} + \xi_i(t).
\label{eq:momentum_change}
\end{align}
The terms $\xi_i(t)$ account for the momentum kicks due to the spontaneous emission of a photon. Whenever a jump from an excited state $| j \rangle$ to the ground state occurs in the Monte Carlo trajectory, we add the recoil momentum $ k_j$ with a random direction to the particle's momentum vector. In our case, we project the three dimensional random momentum vector with length $ k_j$ along the axis determined by the lasers wavevectors. In particular, we have
\begin{align}
\xi_i(t) =   s_i(t) \sum_{j = \{2,3 \} } k_j \delta(t - t_j^{\mathrm{rec}}),
\end{align}
where $s_i(t)$ is the $i$-th component of a random, three-dimensional unit vector. Furthermore, $t_j^{\mathrm{rec}}$ denotes any point in time at which a jump from the $j$-th level occurs. The dissipative processes are the same as described before in equations \eqref{eq:liouvillian_decay} and \eqref{eq:dephasing}.
 
\begin{figure}
\centering
\centering\includegraphics[width=\textwidth]{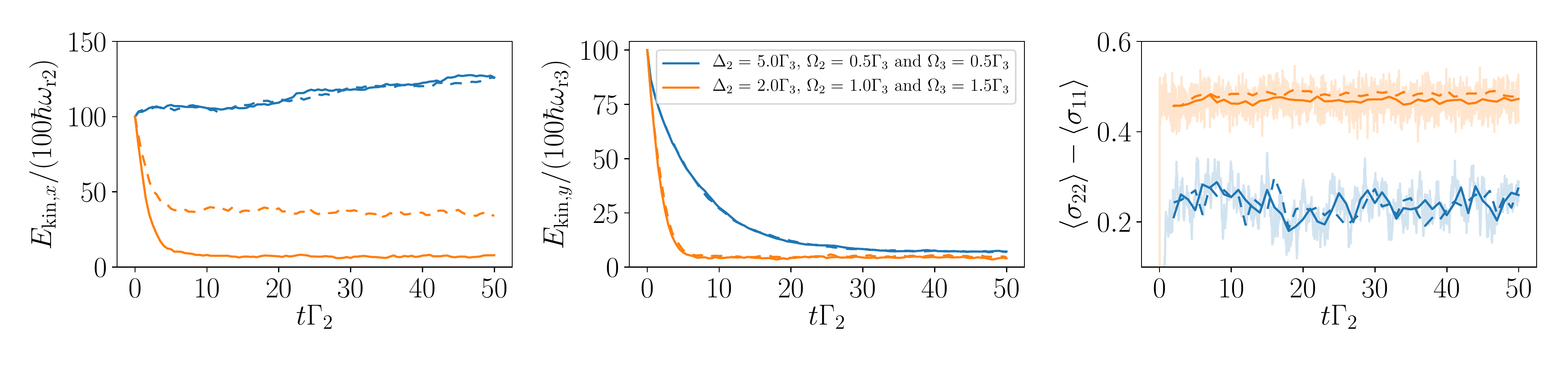}
\caption{\emph{Particle motion}. The solid lines are for $\nu = \Gamma_2$ and the dashed lines for $\nu = 10 \Gamma_2$. We chose $\Delta_3 = -\Gamma_3$. For $\langle \sigma_{22} \rangle - \langle \sigma_{11} \rangle$ we averaged 50 data points to increase the visibility, the bright fast fluctuating lines are the non-averaged for $\nu = 1.0 \Gamma_2$. The atom starts with a momentum of $p_i(0) = 100 k_i$ at the pump field anti-nodes in the ground state. We average 500 MCWF trajectories and the legend is for all three plots.}
\label{fig:cooling_examples}
\end{figure}

In \fref{fig:cooling_examples} we plot the time evolution of the particle's kinetic energy and of the population inversion. For the lasing parameters used before (blue lines) we see that the pump on the lasing transition slowly heats the particles. However, the pump on the broader transition cools them quite well. The lasing transition is still inverted, but the population inversion is significantly decreased. The main reason for this is that, on average, the particle feels a much weaker pump field, since they are not always located at the field maxima. This could be circumvented by simply increasing the laser power. The second set of parameters (orange lines) shows that it is possible to achieve much better cooling on both transitions, and also a larger population inversion. Specifically, the cooling rate is much larger. Unfortunately, these parameters are not suitable for lasing: the comparably small detuning would lead to substantial coherent scattering of the driving laser into the cavity.

Additionally, for the parameters where lasing works well, we find that the finite linewidth of the driving laser rarely affects the particle motion. In the case of optimal cooling, however, the final kinetic energy as well as the cooling time scale is significantly increased. Note also, that the final temperature along the $\textbf{k}_3$-axis is on the order of the Doppler temperature of the broader transition [$k_\mathrm{B} T \approx  \Gamma_3/2$ (blue) and $k_\mathrm{B} T \approx 3 \Gamma_3/10$ (orange)]. The Doppler broadening $\Delta\omega_D$ corresponding to this temperature $k_B T \approx \Gamma_3/2$ on the narrower transition is approximately
\begin{equation}
\Delta\omega_D = \frac{\omega_2}{c} \sqrt{\frac{8 k_B T ln(2)}{m}} \approx 120 \Gamma_2,
\end{equation}
with $\omega_2 = 2\pi \times 435 \text{THz}$ and $m = 87u$. Since this Doppler broadening is approximately a factor of 5 smaller than the power broadened gain ($614 \Gamma_2$, see \fref{fig:atom_spectrum}), the finite temperature does not significantly affect the lasing.

Note that $\Delta_2<0$ and $\Delta_3>0$ is required to achieve cooling. This is because the atoms are inverted on the transition $| 1 \rangle \leftrightarrow | 2 \rangle$, but not on the transition $| 1 \rangle \leftrightarrow | 3 \rangle$ \cite{Ritsch2013cold}. Furthermore, we also want to mention here that for strong pumping and far blue-detuned lasers it is possible that atoms in the ground state get cooled and trapped at the field nodes (high-intensity Sisyphus cooling) \cite{cohen2011advances}. We were able to observe this for some specific parameters on the narrow transition. For the motion of ${}^{174}\text{Yb}$, as mentioned in section \ref{sec:pump_mech}, we almost always obtained heating or trapping at the field nodes.

\section{Conclusions}

We have shown that continuous lasing on a narrow atomic transition can be implemented in a V-level configuration, when a second closed broad transition sharing the same ground state is available. Using two strong and sufficiently detuned coherent driving lasers lead to almost perfect inversion on the narrow transition. The entire mechanism does not rely on a direct decay channel into the excited state with a narrow line, which makes it distinctly different from previously considered lasing setups. When the inverted transition is coupled to an optical resonator, the system starts to lase once a certain threshold number of atoms is passed. The spectral properties of the output laser light exhibit a linewidth that can be well below the natural linewidth of the narrow transition. Notice, that despite the fact that the natural linewidth of the lasing transition is smaller than the cavity linewidth by far, the power broadening induced by the strong pump light means that we are effectively in the good-cavity regime. Furthermore, we found optimal lasing for many atoms and, accordingly, many photons in the cavity. The coherence is thus stored in the cavity field rather than the atomic dipoles. Hence, the lasing setup we consider is more similar to a conventional laser rather than a superradiant one. Yet, the spectral linewidth of the laser can be extremely small and is effectively determined by the natural linewidth of the atom. Finally, we have shown that the overall promising properties of such a laser are conserved even when considering pump lasers that are broad and induce strong dephasing.

Interestingly lasing has been found recently in a closely related setup using Ytterbium atoms \cite{Gothe2019Continuous}. However in this case it was identified as Raman lasing \citep{vrijsen2011raman, bohnet2012steady} which occurs in a very different operation regime and inherits the pump laser linewidth. While our configuration finally does not meet all criteria for a clock laser it can be seen as a major experimental step in this direction.         

In order to estimate heating effects induced by the strong driving beams, we considered the motion of an atom subjected to the inversion scheme. We found that the kinetic energy is limited by the Doppler temperature of the broad transition. Therefore, we conclude that the system should not exhibit instabilities due to heating.

\section{Acknowledgement}
We thank S. Sch\"affer, M. Tang and G. Kazakov for helpful discussions.
This project has received funding from the European Union’s Horizon 2020 research and innovation programme under grant agreement No 820404 (iqClock) (C. H., D. P. and H. R.). 

Numerical simulations were performed with the open source framework QuantumOptics.jl \cite{kramer2018quantumoptics}. The graphs were produced using the open source plotting library Matplotlib~\cite{hunter2007matplotlib}.

\bibliography{VLevelRef}

\begin{thebibliography}{37}%
\makeatletter
\providecommand \@ifxundefined [1]{%
 \@ifx{#1\undefined}
}%
\providecommand \@ifnum [1]{%
 \ifnum #1\expandafter \@firstoftwo
 \else \expandafter \@secondoftwo
 \fi
}%
\providecommand \@ifx [1]{%
 \ifx #1\expandafter \@firstoftwo
 \else \expandafter \@secondoftwo
 \fi
}%
\providecommand \natexlab [1]{#1}%
\providecommand \enquote  [1]{``#1''}%
\providecommand \bibnamefont  [1]{#1}%
\providecommand \bibfnamefont [1]{#1}%
\providecommand \citenamefont [1]{#1}%
\providecommand \href@noop [0]{\@secondoftwo}%
\providecommand \href [0]{\begingroup \@sanitize@url \@href}%
\providecommand \@href[1]{\@@startlink{#1}\@@href}%
\providecommand \@@href[1]{\endgroup#1\@@endlink}%
\providecommand \@sanitize@url [0]{\catcode `\\12\catcode `\$12\catcode
  `\&12\catcode `\#12\catcode `\^12\catcode `\_12\catcode `\%12\relax}%
\providecommand \@@startlink[1]{}%
\providecommand \@@endlink[0]{}%
\providecommand \url  [0]{\begingroup\@sanitize@url \@url }%
\providecommand \@url [1]{\endgroup\@href {#1}{\urlprefix }}%
\providecommand \urlprefix  [0]{URL }%
\providecommand \Eprint [0]{\href }%
\providecommand \doibase [0]{https://doi.org/}%
\providecommand \selectlanguage [0]{\@gobble}%
\providecommand \bibinfo  [0]{\@secondoftwo}%
\providecommand \bibfield  [0]{\@secondoftwo}%
\providecommand \translation [1]{[#1]}%
\providecommand \BibitemOpen [0]{}%
\providecommand \bibitemStop [0]{}%
\providecommand \bibitemNoStop [0]{.\EOS\space}%
\providecommand \EOS [0]{\spacefactor3000\relax}%
\providecommand \BibitemShut  [1]{\csname bibitem#1\endcsname}%
\let\auto@bib@innerbib\@empty
\bibitem [{\citenamefont {Schawlow}\ and\ \citenamefont
  {Townes}(1958)}]{schawlow1958infrared}%
  \BibitemOpen
  \bibfield  {author} {\bibinfo {author} {\bibfnamefont {A.~L.}\ \bibnamefont
  {Schawlow}}\ and\ \bibinfo {author} {\bibfnamefont {C.~H.}\ \bibnamefont
  {Townes}},\ }\href@noop {} {\bibfield  {journal} {\bibinfo  {journal}
  {Physical Review}\ }\textbf {\bibinfo {volume} {112}},\ \bibinfo {pages}
  {1940} (\bibinfo {year} {1958})}\BibitemShut {NoStop}%
\bibitem [{\citenamefont {Strelnitski}\ \emph {et~al.}(1995)\citenamefont
  {Strelnitski}, \citenamefont {Ponomarev},\ and\ \citenamefont
  {Smith}}]{strelnitski1995hydrogen}%
  \BibitemOpen
  \bibfield  {author} {\bibinfo {author} {\bibfnamefont {V.~S.}\ \bibnamefont
  {Strelnitski}}, \bibinfo {author} {\bibfnamefont {V.~O.}\ \bibnamefont
  {Ponomarev}},\ and\ \bibinfo {author} {\bibfnamefont {H.~A.}\ \bibnamefont
  {Smith}},\ }\href@noop {} {\bibfield  {journal} {\bibinfo  {journal} {arXiv
  preprint astro-ph/9511118}\ } (\bibinfo {year} {1995})}\BibitemShut {NoStop}%
\bibitem [{\citenamefont {Goldenberg}\ \emph {et~al.}(1960)\citenamefont
  {Goldenberg}, \citenamefont {Kleppner},\ and\ \citenamefont
  {Ramsey}}]{goldenberg1960atomic}%
  \BibitemOpen
  \bibfield  {author} {\bibinfo {author} {\bibfnamefont {H.~M.}\ \bibnamefont
  {Goldenberg}}, \bibinfo {author} {\bibfnamefont {D.}~\bibnamefont
  {Kleppner}},\ and\ \bibinfo {author} {\bibfnamefont {N.~F.}\ \bibnamefont
  {Ramsey}},\ }\href {https://doi.org/10.1103/PhysRevLett.5.361} {\bibfield
  {journal} {\bibinfo  {journal} {Phys. Rev. Lett.}\ }\textbf {\bibinfo
  {volume} {5}},\ \bibinfo {pages} {361} (\bibinfo {year} {1960})}\BibitemShut
  {NoStop}%
\bibitem [{\citenamefont {Haake}\ \emph {et~al.}(1993)\citenamefont {Haake},
  \citenamefont {Kolobov}, \citenamefont {Fabre}, \citenamefont {Giacobino},\
  and\ \citenamefont {Reynaud}}]{Haake1993superradiant}%
  \BibitemOpen
  \bibfield  {author} {\bibinfo {author} {\bibfnamefont {F.}~\bibnamefont
  {Haake}}, \bibinfo {author} {\bibfnamefont {M.~I.}\ \bibnamefont {Kolobov}},
  \bibinfo {author} {\bibfnamefont {C.}~\bibnamefont {Fabre}}, \bibinfo
  {author} {\bibfnamefont {E.}~\bibnamefont {Giacobino}},\ and\ \bibinfo
  {author} {\bibfnamefont {S.}~\bibnamefont {Reynaud}},\ }\href
  {https://doi.org/10.1103/PhysRevLett.71.995} {\bibfield  {journal} {\bibinfo
  {journal} {Phys. Rev. Lett.}\ }\textbf {\bibinfo {volume} {71}},\ \bibinfo
  {pages} {995} (\bibinfo {year} {1993})}\BibitemShut {NoStop}%
\bibitem [{\citenamefont {Meiser}\ \emph {et~al.}(2009)\citenamefont {Meiser},
  \citenamefont {Ye}, \citenamefont {Carlson},\ and\ \citenamefont
  {Holland}}]{meiser2009prospects}%
  \BibitemOpen
  \bibfield  {author} {\bibinfo {author} {\bibfnamefont {D.}~\bibnamefont
  {Meiser}}, \bibinfo {author} {\bibfnamefont {J.}~\bibnamefont {Ye}}, \bibinfo
  {author} {\bibfnamefont {D.}~\bibnamefont {Carlson}},\ and\ \bibinfo {author}
  {\bibfnamefont {M.}~\bibnamefont {Holland}},\ }\href@noop {} {\bibfield
  {journal} {\bibinfo  {journal} {Physical review letters}\ }\textbf {\bibinfo
  {volume} {102}},\ \bibinfo {pages} {163601} (\bibinfo {year}
  {2009})}\BibitemShut {NoStop}%
\bibitem [{\citenamefont {Norcia}\ and\ \citenamefont
  {Thompson}(2016)}]{norcia2016cold}%
  \BibitemOpen
  \bibfield  {author} {\bibinfo {author} {\bibfnamefont {M.~A.}\ \bibnamefont
  {Norcia}}\ and\ \bibinfo {author} {\bibfnamefont {J.~K.}\ \bibnamefont
  {Thompson}},\ }\href@noop {} {\bibfield  {journal} {\bibinfo  {journal}
  {Physical Review X}\ }\textbf {\bibinfo {volume} {6}},\ \bibinfo {pages}
  {011025} (\bibinfo {year} {2016})}\BibitemShut {NoStop}%
\bibitem [{\citenamefont {Norcia}\ \emph {et~al.}(2018)\citenamefont {Norcia},
  \citenamefont {Cline}, \citenamefont {Muniz}, \citenamefont {Robinson},
  \citenamefont {Hutson}, \citenamefont {Goban}, \citenamefont {Marti},
  \citenamefont {Ye},\ and\ \citenamefont {Thompson}}]{norcia2018frequency}%
  \BibitemOpen
  \bibfield  {author} {\bibinfo {author} {\bibfnamefont {M.~A.}\ \bibnamefont
  {Norcia}}, \bibinfo {author} {\bibfnamefont {J.~R.}\ \bibnamefont {Cline}},
  \bibinfo {author} {\bibfnamefont {J.~A.}\ \bibnamefont {Muniz}}, \bibinfo
  {author} {\bibfnamefont {J.~M.}\ \bibnamefont {Robinson}}, \bibinfo {author}
  {\bibfnamefont {R.~B.}\ \bibnamefont {Hutson}}, \bibinfo {author}
  {\bibfnamefont {A.}~\bibnamefont {Goban}}, \bibinfo {author} {\bibfnamefont
  {G.~E.}\ \bibnamefont {Marti}}, \bibinfo {author} {\bibfnamefont
  {J.}~\bibnamefont {Ye}},\ and\ \bibinfo {author} {\bibfnamefont {J.~K.}\
  \bibnamefont {Thompson}},\ }\href@noop {} {\bibfield  {journal} {\bibinfo
  {journal} {Physical Review X}\ }\textbf {\bibinfo {volume} {8}},\ \bibinfo
  {pages} {021036} (\bibinfo {year} {2018})}\BibitemShut {NoStop}%
\bibitem [{\citenamefont {Laske}\ \emph {et~al.}(2019)\citenamefont {Laske},
  \citenamefont {Winter},\ and\ \citenamefont {Hemmerich}}]{laske2019pulse}%
  \BibitemOpen
  \bibfield  {author} {\bibinfo {author} {\bibfnamefont {T.}~\bibnamefont
  {Laske}}, \bibinfo {author} {\bibfnamefont {H.}~\bibnamefont {Winter}},\ and\
  \bibinfo {author} {\bibfnamefont {A.}~\bibnamefont {Hemmerich}},\ }\href@noop
  {} {\bibfield  {journal} {\bibinfo  {journal} {Physical Review Letters}\
  }\textbf {\bibinfo {volume} {123}},\ \bibinfo {pages} {103601} (\bibinfo
  {year} {2019})}\BibitemShut {NoStop}%
\bibitem [{\citenamefont {Sch{\"a}ffer}\ \emph {et~al.}(2020)\citenamefont
  {Sch{\"a}ffer}, \citenamefont {Tang}, \citenamefont {Henriksen},
  \citenamefont {J{\o}rgensen}, \citenamefont {Christensen},\ and\
  \citenamefont {Thomsen}}]{schaffer2020lasing}%
  \BibitemOpen
  \bibfield  {author} {\bibinfo {author} {\bibfnamefont {S.~A.}\ \bibnamefont
  {Sch{\"a}ffer}}, \bibinfo {author} {\bibfnamefont {M.}~\bibnamefont {Tang}},
  \bibinfo {author} {\bibfnamefont {M.~R.}\ \bibnamefont {Henriksen}}, \bibinfo
  {author} {\bibfnamefont {A.~A.}\ \bibnamefont {J{\o}rgensen}}, \bibinfo
  {author} {\bibfnamefont {B.~T.}\ \bibnamefont {Christensen}},\ and\ \bibinfo
  {author} {\bibfnamefont {J.~W.}\ \bibnamefont {Thomsen}},\ }\href@noop {}
  {\bibfield  {journal} {\bibinfo  {journal} {Physical Review a}\ }\textbf
  {\bibinfo {volume} {101}},\ \bibinfo {pages} {013819} (\bibinfo {year}
  {2020})}\BibitemShut {NoStop}%
\bibitem [{\citenamefont {Gogyan}\ \emph {et~al.}(2020)\citenamefont {Gogyan},
  \citenamefont {Kazakov}, \citenamefont {Bober},\ and\ \citenamefont
  {Zawada}}]{Gogyan2020characterisation}%
  \BibitemOpen
  \bibfield  {author} {\bibinfo {author} {\bibfnamefont {A.}~\bibnamefont
  {Gogyan}}, \bibinfo {author} {\bibfnamefont {G.}~\bibnamefont {Kazakov}},
  \bibinfo {author} {\bibfnamefont {M.}~\bibnamefont {Bober}},\ and\ \bibinfo
  {author} {\bibfnamefont {M.}~\bibnamefont {Zawada}},\ }\href
  {https://doi.org/10.1364/OE.381991} {\bibfield  {journal} {\bibinfo
  {journal} {Opt. Express}\ }\textbf {\bibinfo {volume} {28}},\ \bibinfo
  {pages} {6881} (\bibinfo {year} {2020})}\BibitemShut {NoStop}%
\bibitem [{\citenamefont {Bohnet}\ \emph {et~al.}(2012)\citenamefont {Bohnet},
  \citenamefont {Chen}, \citenamefont {Weiner}, \citenamefont {Meiser},
  \citenamefont {Holland},\ and\ \citenamefont {Thompson}}]{bohnet2012steady}%
  \BibitemOpen
  \bibfield  {author} {\bibinfo {author} {\bibfnamefont {J.~G.}\ \bibnamefont
  {Bohnet}}, \bibinfo {author} {\bibfnamefont {Z.}~\bibnamefont {Chen}},
  \bibinfo {author} {\bibfnamefont {J.~M.}\ \bibnamefont {Weiner}}, \bibinfo
  {author} {\bibfnamefont {D.}~\bibnamefont {Meiser}}, \bibinfo {author}
  {\bibfnamefont {M.~J.}\ \bibnamefont {Holland}},\ and\ \bibinfo {author}
  {\bibfnamefont {J.~K.}\ \bibnamefont {Thompson}},\ }\href@noop {} {\bibfield
  {journal} {\bibinfo  {journal} {Nature}\ }\textbf {\bibinfo {volume} {484}},\
  \bibinfo {pages} {78} (\bibinfo {year} {2012})}\BibitemShut {NoStop}%
\bibitem [{\citenamefont {Maier}\ \emph {et~al.}(2014)\citenamefont {Maier},
  \citenamefont {Kraemer}, \citenamefont {Ostermann},\ and\ \citenamefont
  {Ritsch}}]{maier2014superradiant}%
  \BibitemOpen
  \bibfield  {author} {\bibinfo {author} {\bibfnamefont {T.}~\bibnamefont
  {Maier}}, \bibinfo {author} {\bibfnamefont {S.}~\bibnamefont {Kraemer}},
  \bibinfo {author} {\bibfnamefont {L.}~\bibnamefont {Ostermann}},\ and\
  \bibinfo {author} {\bibfnamefont {H.}~\bibnamefont {Ritsch}},\ }\href@noop {}
  {\bibfield  {journal} {\bibinfo  {journal} {Optics express}\ }\textbf
  {\bibinfo {volume} {22}},\ \bibinfo {pages} {13269} (\bibinfo {year}
  {2014})}\BibitemShut {NoStop}%
\bibitem [{\citenamefont {Norcia}\ \emph {et~al.}(2016)\citenamefont {Norcia},
  \citenamefont {Winchester}, \citenamefont {Cline},\ and\ \citenamefont
  {Thompson}}]{norcia2016superradiance}%
  \BibitemOpen
  \bibfield  {author} {\bibinfo {author} {\bibfnamefont {M.~A.}\ \bibnamefont
  {Norcia}}, \bibinfo {author} {\bibfnamefont {M.~N.}\ \bibnamefont
  {Winchester}}, \bibinfo {author} {\bibfnamefont {J.~R.}\ \bibnamefont
  {Cline}},\ and\ \bibinfo {author} {\bibfnamefont {J.~K.}\ \bibnamefont
  {Thompson}},\ }\href@noop {} {\bibfield  {journal} {\bibinfo  {journal}
  {Science advances}\ }\textbf {\bibinfo {volume} {2}},\ \bibinfo {pages}
  {e1601231} (\bibinfo {year} {2016})}\BibitemShut {NoStop}%
\bibitem [{\citenamefont {Numata}\ \emph {et~al.}(2004)\citenamefont {Numata},
  \citenamefont {Kemery},\ and\ \citenamefont {Camp}}]{numata2004thermal}%
  \BibitemOpen
  \bibfield  {author} {\bibinfo {author} {\bibfnamefont {K.}~\bibnamefont
  {Numata}}, \bibinfo {author} {\bibfnamefont {A.}~\bibnamefont {Kemery}},\
  and\ \bibinfo {author} {\bibfnamefont {J.}~\bibnamefont {Camp}},\ }\href@noop
  {} {\bibfield  {journal} {\bibinfo  {journal} {Physical review letters}\
  }\textbf {\bibinfo {volume} {93}},\ \bibinfo {pages} {250602} (\bibinfo
  {year} {2004})}\BibitemShut {NoStop}%
\bibitem [{\citenamefont {Chen}(2009)}]{Chen2009active}%
  \BibitemOpen
  \bibfield  {author} {\bibinfo {author} {\bibfnamefont {J.}~\bibnamefont
  {Chen}},\ }\href {https://doi.org/10.1007/s11434-009-0073-y} {\bibfield
  {journal} {\bibinfo  {journal} {Chinese Science Bulletin}\ }\textbf {\bibinfo
  {volume} {54}},\ \bibinfo {pages} {348} (\bibinfo {year} {2009})}\BibitemShut
  {NoStop}%
\bibitem [{\citenamefont {Kazakov}\ and\ \citenamefont
  {Schumm}(2014)}]{Kazakov2014active}%
  \BibitemOpen
  \bibfield  {author} {\bibinfo {author} {\bibfnamefont {G.~A.}\ \bibnamefont
  {Kazakov}}\ and\ \bibinfo {author} {\bibfnamefont {T.}~\bibnamefont
  {Schumm}},\ }in\ \href@noop {} {\emph {\bibinfo {booktitle} {2014 European
  Frequency and Time Forum (EFTF)}}}\ (\bibinfo {organization} {IEEE},\
  \bibinfo {year} {2014})\ pp.\ \bibinfo {pages} {411--414}\BibitemShut
  {NoStop}%
\bibitem [{\citenamefont {Chen}\ \emph {et~al.}(2019)\citenamefont {Chen},
  \citenamefont {Bennetts}, \citenamefont {Escudero}, \citenamefont
  {Pasquiou},\ and\ \citenamefont {Schreck}}]{Chen2019continuous}%
  \BibitemOpen
  \bibfield  {author} {\bibinfo {author} {\bibfnamefont {C.-C.}\ \bibnamefont
  {Chen}}, \bibinfo {author} {\bibfnamefont {S.}~\bibnamefont {Bennetts}},
  \bibinfo {author} {\bibfnamefont {R.~G.}\ \bibnamefont {Escudero}}, \bibinfo
  {author} {\bibfnamefont {B.}~\bibnamefont {Pasquiou}},\ and\ \bibinfo
  {author} {\bibfnamefont {F.}~\bibnamefont {Schreck}},\ }\href
  {https://doi.org/10.1103/PhysRevApplied.12.044014} {\bibfield  {journal}
  {\bibinfo  {journal} {Phys. Rev. Applied}\ }\textbf {\bibinfo {volume}
  {12}},\ \bibinfo {pages} {044014} (\bibinfo {year} {2019})}\BibitemShut
  {NoStop}%
\bibitem [{\citenamefont {Meduri}\ \emph {et~al.}(1993)\citenamefont {Meduri},
  \citenamefont {Wilson}, \citenamefont {Sellin},\ and\ \citenamefont
  {Mossberg}}]{Meduri1993dynamical}%
  \BibitemOpen
  \bibfield  {author} {\bibinfo {author} {\bibfnamefont {K.~K.}\ \bibnamefont
  {Meduri}}, \bibinfo {author} {\bibfnamefont {G.~A.}\ \bibnamefont {Wilson}},
  \bibinfo {author} {\bibfnamefont {P.~B.}\ \bibnamefont {Sellin}},\ and\
  \bibinfo {author} {\bibfnamefont {T.~W.}\ \bibnamefont {Mossberg}},\ }\href
  {https://doi.org/10.1103/PhysRevLett.71.4311} {\bibfield  {journal} {\bibinfo
   {journal} {Phys. Rev. Lett.}\ }\textbf {\bibinfo {volume} {71}},\ \bibinfo
  {pages} {4311} (\bibinfo {year} {1993})}\BibitemShut {NoStop}%
\bibitem [{\citenamefont {Gothe}\ \emph {et~al.}(2019)\citenamefont {Gothe},
  \citenamefont {Sholokhov}, \citenamefont {Breunig}, \citenamefont {Steinel},\
  and\ \citenamefont {Eschner}}]{Gothe2019Continuous}%
  \BibitemOpen
  \bibfield  {author} {\bibinfo {author} {\bibfnamefont {H.}~\bibnamefont
  {Gothe}}, \bibinfo {author} {\bibfnamefont {D.}~\bibnamefont {Sholokhov}},
  \bibinfo {author} {\bibfnamefont {A.}~\bibnamefont {Breunig}}, \bibinfo
  {author} {\bibfnamefont {M.}~\bibnamefont {Steinel}},\ and\ \bibinfo {author}
  {\bibfnamefont {J.}~\bibnamefont {Eschner}},\ }\href
  {https://doi.org/10.1103/PhysRevA.99.013415} {\bibfield  {journal} {\bibinfo
  {journal} {Phys. Rev. A}\ }\textbf {\bibinfo {volume} {99}},\ \bibinfo
  {pages} {013415} (\bibinfo {year} {2019})}\BibitemShut {NoStop}%
\bibitem [{\citenamefont {Plankensteiner}\ \emph {et~al.}(2016)\citenamefont
  {Plankensteiner}, \citenamefont {Schachenmayer}, \citenamefont {Ritsch},\
  and\ \citenamefont {Genes}}]{Plankensteiner2016laser_noise}%
  \BibitemOpen
  \bibfield  {author} {\bibinfo {author} {\bibfnamefont {D.}~\bibnamefont
  {Plankensteiner}}, \bibinfo {author} {\bibfnamefont {J.}~\bibnamefont
  {Schachenmayer}}, \bibinfo {author} {\bibfnamefont {H.}~\bibnamefont
  {Ritsch}},\ and\ \bibinfo {author} {\bibfnamefont {C.}~\bibnamefont
  {Genes}},\ }\href {https://doi.org/10.1088/0953-4075/49/24/245501} {\bibfield
   {journal} {\bibinfo  {journal} {Journal of Physics B: Atomic, Molecular and
  Optical Physics}\ }\textbf {\bibinfo {volume} {49}},\ \bibinfo {pages}
  {245501} (\bibinfo {year} {2016})}\BibitemShut {NoStop}%
\bibitem [{\citenamefont {Dorner}(2012)}]{Dorner2012quantum_frequency}%
  \BibitemOpen
  \bibfield  {author} {\bibinfo {author} {\bibfnamefont {U.}~\bibnamefont
  {Dorner}},\ }\href {https://doi.org/10.1088/1367-2630/14/4/043011} {\bibfield
   {journal} {\bibinfo  {journal} {New Journal of Physics}\ }\textbf {\bibinfo
  {volume} {14}},\ \bibinfo {pages} {043011} (\bibinfo {year}
  {2012})}\BibitemShut {NoStop}%
\bibitem [{\citenamefont {Kubo}(1962)}]{Kubo1962generalized}%
  \BibitemOpen
  \bibfield  {author} {\bibinfo {author} {\bibfnamefont {R.}~\bibnamefont
  {Kubo}},\ }\href@noop {} {\bibfield  {journal} {\bibinfo  {journal} {Journal
  of the Physical Society of Japan}\ }\textbf {\bibinfo {volume} {17}},\
  \bibinfo {pages} {1100} (\bibinfo {year} {1962})}\BibitemShut {NoStop}%
\bibitem [{\citenamefont {Puri}(2001)}]{puri2001mathematical}%
  \BibitemOpen
  \bibfield  {author} {\bibinfo {author} {\bibfnamefont {R.~R.}\ \bibnamefont
  {Puri}},\ }\href@noop {} {\emph {\bibinfo {title} {Mathematical methods of
  quantum optics}}}\ (\bibinfo  {publisher} {Springer Berlin Heidelberg},\
  \bibinfo {year} {2001})\BibitemShut {NoStop}%
\bibitem [{\citenamefont {Carmichael}(2013)}]{carmichael2013statistical}%
  \BibitemOpen
  \bibfield  {author} {\bibinfo {author} {\bibfnamefont {H.~J.}\ \bibnamefont
  {Carmichael}},\ }\href@noop {} {\emph {\bibinfo {title} {Statistical methods
  in quantum optics 1: master equations and Fokker-Planck equations}}}\
  (\bibinfo  {publisher} {Springer Science \& Business Media},\ \bibinfo {year}
  {2013})\BibitemShut {NoStop}%
\bibitem [{\citenamefont {Avan}\ and\ \citenamefont
  {Cohen-Tannoudji}(1977)}]{avan1977two}%
  \BibitemOpen
  \bibfield  {author} {\bibinfo {author} {\bibfnamefont {P.}~\bibnamefont
  {Avan}}\ and\ \bibinfo {author} {\bibfnamefont {C.}~\bibnamefont
  {Cohen-Tannoudji}},\ }\href@noop {} {\bibfield  {journal} {\bibinfo
  {journal} {Journal of Physics B: Atomic and Molecular Physics}\ }\textbf
  {\bibinfo {volume} {10}},\ \bibinfo {pages} {155} (\bibinfo {year}
  {1977})}\BibitemShut {NoStop}%
\bibitem [{\citenamefont {Henschel}\ \emph {et~al.}(2010)\citenamefont
  {Henschel}, \citenamefont {Majer}, \citenamefont {Schmiedmayer},\ and\
  \citenamefont {Ritsch}}]{henschel2010cavity}%
  \BibitemOpen
  \bibfield  {author} {\bibinfo {author} {\bibfnamefont {K.}~\bibnamefont
  {Henschel}}, \bibinfo {author} {\bibfnamefont {J.}~\bibnamefont {Majer}},
  \bibinfo {author} {\bibfnamefont {J.}~\bibnamefont {Schmiedmayer}},\ and\
  \bibinfo {author} {\bibfnamefont {H.}~\bibnamefont {Ritsch}},\ }\href@noop {}
  {\bibfield  {journal} {\bibinfo  {journal} {Physical Review A}\ }\textbf
  {\bibinfo {volume} {82}},\ \bibinfo {pages} {033810} (\bibinfo {year}
  {2010})}\BibitemShut {NoStop}%
\bibitem [{\citenamefont {Dum}\ \emph {et~al.}(1992)\citenamefont {Dum},
  \citenamefont {Zoller},\ and\ \citenamefont {Ritsch}}]{Dum1992montecarlo}%
  \BibitemOpen
  \bibfield  {author} {\bibinfo {author} {\bibfnamefont {R.}~\bibnamefont
  {Dum}}, \bibinfo {author} {\bibfnamefont {P.}~\bibnamefont {Zoller}},\ and\
  \bibinfo {author} {\bibfnamefont {H.}~\bibnamefont {Ritsch}},\ }\href
  {https://doi.org/10.1103/PhysRevA.45.4879} {\bibfield  {journal} {\bibinfo
  {journal} {Phys. Rev. A}\ }\textbf {\bibinfo {volume} {45}},\ \bibinfo
  {pages} {4879} (\bibinfo {year} {1992})}\BibitemShut {NoStop}%
\bibitem [{\citenamefont {M{\o}lmer}\ \emph {et~al.}(1993)\citenamefont
  {M{\o}lmer}, \citenamefont {Castin},\ and\ \citenamefont
  {Dalibard}}]{Molmer1993MonteCarlo}%
  \BibitemOpen
  \bibfield  {author} {\bibinfo {author} {\bibfnamefont {K.}~\bibnamefont
  {M{\o}lmer}}, \bibinfo {author} {\bibfnamefont {Y.}~\bibnamefont {Castin}},\
  and\ \bibinfo {author} {\bibfnamefont {J.}~\bibnamefont {Dalibard}},\ }\href
  {https://doi.org/10.1364/JOSAB.10.000524} {\bibfield  {journal} {\bibinfo
  {journal} {J. Opt. Soc. Am. B}\ }\textbf {\bibinfo {volume} {10}},\ \bibinfo
  {pages} {524} (\bibinfo {year} {1993})}\BibitemShut {NoStop}%
\bibitem [{\citenamefont {Salzburger}\ and\ \citenamefont
  {Ritsch}(2004)}]{salzburger2004atomic}%
  \BibitemOpen
  \bibfield  {author} {\bibinfo {author} {\bibfnamefont {T.}~\bibnamefont
  {Salzburger}}\ and\ \bibinfo {author} {\bibfnamefont {H.}~\bibnamefont
  {Ritsch}},\ }\href@noop {} {\bibfield  {journal} {\bibinfo  {journal}
  {Physical review letters}\ }\textbf {\bibinfo {volume} {93}},\ \bibinfo
  {pages} {063002} (\bibinfo {year} {2004})}\BibitemShut {NoStop}%
\bibitem [{\citenamefont {Hotter}\ \emph {et~al.}(2019)\citenamefont {Hotter},
  \citenamefont {Plankensteiner}, \citenamefont {Ostermann},\ and\
  \citenamefont {Ritsch}}]{hotter2019superradiant}%
  \BibitemOpen
  \bibfield  {author} {\bibinfo {author} {\bibfnamefont {C.}~\bibnamefont
  {Hotter}}, \bibinfo {author} {\bibfnamefont {D.}~\bibnamefont
  {Plankensteiner}}, \bibinfo {author} {\bibfnamefont {L.}~\bibnamefont
  {Ostermann}},\ and\ \bibinfo {author} {\bibfnamefont {H.}~\bibnamefont
  {Ritsch}},\ }\href@noop {} {\bibfield  {journal} {\bibinfo  {journal} {Optics
  express}\ }\textbf {\bibinfo {volume} {27}},\ \bibinfo {pages} {31193}
  (\bibinfo {year} {2019})}\BibitemShut {NoStop}%
\bibitem [{\citenamefont {Ritsch}\ \emph {et~al.}(2013)\citenamefont {Ritsch},
  \citenamefont {Domokos}, \citenamefont {Brennecke},\ and\ \citenamefont
  {Esslinger}}]{Ritsch2013cold}%
  \BibitemOpen
  \bibfield  {author} {\bibinfo {author} {\bibfnamefont {H.}~\bibnamefont
  {Ritsch}}, \bibinfo {author} {\bibfnamefont {P.}~\bibnamefont {Domokos}},
  \bibinfo {author} {\bibfnamefont {F.}~\bibnamefont {Brennecke}},\ and\
  \bibinfo {author} {\bibfnamefont {T.}~\bibnamefont {Esslinger}},\ }\href
  {https://doi.org/10.1103/RevModPhys.85.553} {\bibfield  {journal} {\bibinfo
  {journal} {Rev. Mod. Phys.}\ }\textbf {\bibinfo {volume} {85}},\ \bibinfo
  {pages} {553} (\bibinfo {year} {2013})}\BibitemShut {NoStop}%
\bibitem [{\citenamefont {Cohen-Tannoudji}\ and\ \citenamefont
  {Gu{\'e}ry-Odelin}(2011)}]{cohen2011advances}%
  \BibitemOpen
  \bibfield  {author} {\bibinfo {author} {\bibfnamefont {C.}~\bibnamefont
  {Cohen-Tannoudji}}\ and\ \bibinfo {author} {\bibfnamefont {D.}~\bibnamefont
  {Gu{\'e}ry-Odelin}},\ }\href {https://books.google.at/books?id=md\_cNwAACAAJ}
  {\emph {\bibinfo {title} {Advances in Atomic Physics: An Overview}}}\
  (\bibinfo  {publisher} {World Scientific},\ \bibinfo {year}
  {2011})\BibitemShut {NoStop}%
\bibitem [{\citenamefont {Vrijsen}\ \emph {et~al.}(2011)\citenamefont
  {Vrijsen}, \citenamefont {Hosten}, \citenamefont {Lee}, \citenamefont
  {Bernon},\ and\ \citenamefont {Kasevich}}]{vrijsen2011raman}%
  \BibitemOpen
  \bibfield  {author} {\bibinfo {author} {\bibfnamefont {G.}~\bibnamefont
  {Vrijsen}}, \bibinfo {author} {\bibfnamefont {O.}~\bibnamefont {Hosten}},
  \bibinfo {author} {\bibfnamefont {J.}~\bibnamefont {Lee}}, \bibinfo {author}
  {\bibfnamefont {S.}~\bibnamefont {Bernon}},\ and\ \bibinfo {author}
  {\bibfnamefont {M.~A.}\ \bibnamefont {Kasevich}},\ }\href@noop {} {\bibfield
  {journal} {\bibinfo  {journal} {Physical review letters}\ }\textbf {\bibinfo
  {volume} {107}},\ \bibinfo {pages} {063904} (\bibinfo {year}
  {2011})}\BibitemShut {NoStop}%
\bibitem [{\citenamefont {Kr{\"a}mer}\ \emph {et~al.}(2018)\citenamefont
  {Kr{\"a}mer}, \citenamefont {Plankensteiner}, \citenamefont {Ostermann},\
  and\ \citenamefont {Ritsch}}]{kramer2018quantumoptics}%
  \BibitemOpen
  \bibfield  {author} {\bibinfo {author} {\bibfnamefont {S.}~\bibnamefont
  {Kr{\"a}mer}}, \bibinfo {author} {\bibfnamefont {D.}~\bibnamefont
  {Plankensteiner}}, \bibinfo {author} {\bibfnamefont {L.}~\bibnamefont
  {Ostermann}},\ and\ \bibinfo {author} {\bibfnamefont {H.}~\bibnamefont
  {Ritsch}},\ }\href {https://doi.org/10.1016/j.cpc.2018.02.004} {\bibfield
  {journal} {\bibinfo  {journal} {Comput. Phys. Commun.}\ }\textbf {\bibinfo
  {volume} {227}},\ \bibinfo {pages} {109} (\bibinfo {year}
  {2018})}\BibitemShut {NoStop}%
\bibitem [{\citenamefont {Hunter}(2007)}]{hunter2007matplotlib}%
  \BibitemOpen
  \bibfield  {author} {\bibinfo {author} {\bibfnamefont {J.~D.}\ \bibnamefont
  {Hunter}},\ }\href@noop {} {\bibfield  {journal} {\bibinfo  {journal}
  {Computing in science \& engineering}\ }\textbf {\bibinfo {volume} {9}},\
  \bibinfo {pages} {90} (\bibinfo {year} {2007})}\BibitemShut {NoStop}%
\bibitem [{\citenamefont {Gardiner}\ \emph {et~al.}(1985)\citenamefont
  {Gardiner} \emph {et~al.}}]{gardiner1985handbook}%
  \BibitemOpen
  \bibfield  {author} {\bibinfo {author} {\bibfnamefont {C.~W.}\ \bibnamefont
  {Gardiner}} \emph {et~al.},\ }\href@noop {} {\emph {\bibinfo {title}
  {Handbook of stochastic methods}}},\ Vol.~\bibinfo {volume} {3}\ (\bibinfo
  {publisher} {springer Berlin},\ \bibinfo {year} {1985})\BibitemShut {NoStop}%
\bibitem [{git()}]{github_qumulants}%
  \BibitemOpen
  \href@noop {} {\bibinfo {title} {{GitHub} repository for the {Qumulants.jl}
  package}},\ \bibinfo {howpublished}
  {\url{https://github.com/david-pl/Qumulants.jl}},\ \bibinfo {note} {accessed:
  \today}\BibitemShut {NoStop}%
\end{thebibliography}%

\newpage

\appendix 
\section{Dephasing due to a finite pump laser linewidth} \label{sec:dephasing_pump}

In this section we show the fundamental steps to derive the dephasing Liouvillian originating from a finite pump laser linewidth for $N$ coherently driven two-level atoms coupled to a cavity \eqref{eq:dephasing2_N}. Using the standard phase diffusion model with a noisy phase $\phi_2(t)$ we have the following Hamiltonian written in the lab frame
\begin{equation}
H_\mathrm{lab} = \omega_{\mathrm{c}} a^{\dagger}a + \sum_{j=1}^N g_j (a^{\dagger} \sigma_{12}^{j} + a \sigma_{21}^{j}) + \sum_{j=1}^N \omega_2^{j} \sigma_{22}^{j} + \sum_{j=1}^N \Omega_2^{j} \big(\sigma_{21}^{j} e^{-i(\omega_{\ell 2}t + \phi_2(t))} + \sigma_{12}^{j}e^{i(\omega_{\ell 2}t + \phi_2(t))} \big) ,
\label{eq:Hamiltonian_lasing_filter_cavity_dephasing}
\end{equation}
where the noise statistics of $\phi_2(t)$ is determined by its derivative, which is assumed to be a white noise frequency fluctuation such that
\begin{equation}
\langle \dot{\phi}_2(t) \dot{\phi}_2(t') \rangle = \nu_2 \delta(t-t') .
\label{eq:white_noise_phase}
\end{equation}
Additionally, we assume here that all atoms experience the same pump laser phase (collective phase noise). We switch into the (instantaneous) rotating frame of the pump laser with the unitary transformation
\begin{equation}
U_I(t) = e^{i(\omega_{\ell 2}t + \phi_2(t))(a^\dagger a + \sum_{j=1}^N \sigma_{22}^j)},
\label{eq:rot_fram}
\end{equation}
in which the Hamiltonian becomes
\begin{equation}
H = -\Delta_{\mathrm{c}} a^{\dagger}a + \sum_{j=1}^N g_j (a^{\dagger} \sigma_{12}^{j} + a \sigma_{21}^{j}) + \sum_{j=1}^N -\Delta_2^{j} \sigma_{22}^{j} + \Omega_2^{j} (\sigma_{21}^{j} + \sigma_{12}^{j}) - \dot{\phi}_2(t) (a^\dagger a + \sum_{j=1}^N \sigma_{22}^j) .
\label{eq:H_rot_frame}
\end{equation}
We distinguish now between the deterministic part of the Hamiltonian $H_D$ and the stochastic part 
\begin{equation}
H_S = - (a^\dagger a + \sum_{j=1}^N \sigma_{22}^j) ,
\label{eq:H_stoch}
\end{equation}
such that $H = H_D + \dot{\phi}_2(t) H_S$. The Heisenberg equation of a system operator $O$ for this Hamiltonian can be written as
\begin{equation}
\text{(S)} \hspace{0.5cm} \frac{d}{dt} O = L_D[O] + \dot{\phi}_2(t) L_S[O],
\label{eq:heisenberg}
\end{equation}
with $L_D[O] = i [H_D , O]$ and $L_S[O] = i [H_S , O]$.
Equation \eqref{eq:heisenberg} needs to be interpreted as a Stratonovich stochastic differential equation [indicated by (S)] , which can be transformed into It\^o form [indicated by (I)] as follows \cite{gardiner1985handbook}:
\begin{equation}
\text{(I)} \hspace{0.5cm} \frac{d}{dt} O = L_D[O] + \frac{1}{2} \nu_2 L_S^2[O] + \dot{\phi}_2(t) L_S[O] ,
\label{eq:ito_heisenberg}
\end{equation}
By averaging equation \eqref{eq:ito_heisenberg} the stochastic part vanishes and we get 
\begin{equation}
\frac{d}{dt} \braket{O}_S = \braket{L_D[O]}_S + \frac{1}{2} \nu_2 \braket{L_S^2[O]}_S .
\label{eq:ito_heisenberg_avg}
\end{equation}
Evaluating $L_S^2[O] = -[H_S,[H_S,O]]$, we find that the fluctuating phase leads to dephasing as described in equation \eqref{eq:dephasing2_N}. Note, that $\braket{ \cdot }_S$ indicates a stochastic average, not a quantum average as in the cumulant expansion.

\section{Derivation of correlation function equations} \label{sec:corr_deriv}
On the one hand, we compute the steady-state expectation values in the instantaneous rotating frame (see \ref{sec:dephasing_pump}). On the other hand, the correlation function for the spectrum has to be computed in a non-fluctuating (coherent) rotating frame \cite{avan1977two}. In this section we show the main procedure to derive the equations. The full set of equations to obtain the correlation function is given in \ref{sec:corr_fct_eqs}. Again, we consider $N$ two-level atoms coupled to a cavity as in \ref{sec:dephasing_pump}, including the third level is straight forward. In the coherent rotating frame with the unitary transformation 
\begin{equation}
U_C(t) = e^{i\omega_{\ell 2}t(a^\dagger a + \sum_{j=1}^N \sigma_{22}^j)} ,
\label{eq:coherent_rot_fram}
\end{equation}
we obtain the time-dependent Hamiltonian
\begin{equation}
\tilde{H} = - \Delta_{\mathrm{c}} \tilde{a}^{\dagger}\tilde{a} + \sum_{j=1}^N g_j (\tilde{a}^{\dagger} \tilde{\sigma}_{12}^{j} + \tilde{a} \tilde{\sigma}_{21}^{j}) - \sum_{j=1}^N \Delta_2^{j} \tilde{\sigma}_{22}^{j} + \sum_{j=1}^N \Omega_2^{j} \big(\tilde{\sigma}_{21}^{j} e^{-i \phi_2(t)} + \tilde{\sigma}_{12}^{j}e^{i \phi_2(t)} \big) .
\label{eq:Hamiltonian_corr}
\end{equation}
The tilde indicates that the operator is in the coherent rotating frame. With cavity decay and individual atomic decay, described by \eqref{eq:cavity_decay} and \eqref{eq:ind_decay_sum}, respectively, we obtain with the quantum regression theorem \cite{carmichael2013statistical} the differential equation for the correlation function
\begin{equation}
\frac{d}{dt} \langle \tilde{a}^\dagger  \tilde{a}_0\rangle  =  i g N \langle \tilde\sigma^{1}_{21}  \tilde{a}_0\rangle - i \Delta_{c} \langle \tilde{a}^\dagger  \tilde{a}_0\rangle  -0.5 \kappa \langle \tilde{a}^\dagger  \tilde{a}_0\rangle 
\label{eq:corr_ada}
\end{equation}
Here and in the the following, we use the notation $ \braket{\tilde{O}(t) \tilde{a}(0)} = \braket{\tilde{O} \tilde{a}_0} $. To calculate \eqref{eq:corr_ada} we also need the equation for $\braket{\tilde{\sigma}_1^{21}(t)\tilde{a}(0)}$ which is given by
\begin{equation}
\begin{aligned}
\frac{d}{dt} \langle \tilde\sigma^{1}_{21}  \tilde{a}_0\rangle  &= -2 i g \left( \langle \tilde{a}^\dagger\rangle  \langle \tilde\sigma^{1}_{22}  \tilde{a}_0\rangle  + \langle \tilde{a}_0\rangle  \langle \tilde{a}^\dagger  \tilde\sigma^{1}_{22}\rangle  + \langle \tilde\sigma^{1}_{22}\rangle  \langle \tilde{a}^\dagger  \tilde{a}_0\rangle  -2 \langle \tilde{a}^\dagger\rangle  \langle \tilde{a}_0\rangle  \langle \tilde\sigma^{1}_{22}\rangle  \right) \\
&+  i g \langle \tilde{a}^\dagger  \tilde{a}_0\rangle  -0.5 \Gamma_{2} \langle \tilde\sigma^{1}_{21}  \tilde{a}_0\rangle  - i \Delta_{2} \langle \tilde\sigma^{1}_{21}  \tilde{a}_0\rangle  +  i \Omega_{2} \langle \tilde{a}_0\rangle e^{ i \phi_2{(t)}} -2 i \Omega_{2} \langle \tilde\sigma^{1}_{22}  \tilde{a}_0\rangle  e^{ i \phi_2{(t)}} .
\end{aligned}
\label{eq:corr_s21a}
\end{equation}
Equation \eqref{eq:corr_s21a} contains some specific averages proportional to $e^{i \phi(t)}$ ($ \braket{\tilde{a}}$ and $\braket{\tilde{\sigma}_{22}^1  \tilde{a}_0}$). To solve this problem we use two properties: first, since we are in steady state, all averages that do not involve $\tilde{a}_0$ do not depend on two different times and hence can be replaced by the respective steady-state values. Second, the change of the steady-state values from the coherent to the instantaneous frame is given by the unitary transformation
\begin{equation}
U_{C \rightarrow I}(t) = e^{i \phi_2(t) (a^\dagger a + \sum_{j=1}^N \sigma_{22}^j)} .
\label{eq:coh_inst_trans}
\end{equation}
We get for example $\braket{\tilde{a}}  = \braket{a} e^{-i \phi_2(t)}$. With this we can use the steady-state values in the instantaneous frame, which we have obtained before. Replacing them, we find that all averages $\braket{\tilde{\sigma}_{22}^1 \tilde{a}_0}$ only occur in combination with the factor $e^{i \phi_2(t)}$. Therefore, we derive the equations of motion for $\braket{\tilde{\sigma}_{22}^1 \tilde{a}_0} e^{i \phi_2(t)}$ rather than $\braket{\tilde{\sigma}_{22}^1 \tilde{a}_0}$ alone. In equation \eqref{eq:corr_sa} we show two examples of such differential equations for averages multiplied with a phase fluctuation term:
\begin{subequations}
\begin{equation}
\begin{aligned}
\frac{d}{dt} \langle \tilde\sigma^{1}_{22}  \tilde{a}_0\rangle e^{ i {\phi_{2}(t)}}  &=  i \dot\phi_{2} \langle \tilde\sigma^{1}_{22}  \tilde{a}_0\rangle  +  i g \left( \langle \tilde{a}^\dagger\rangle  \langle \tilde\sigma^{1}_{12}  \tilde{a}_0\rangle  + \langle \tilde{a}_0\rangle  \langle \tilde{a}^\dagger  \tilde\sigma^{1}_{12}\rangle  + \langle \tilde\sigma^{1}_{12}\rangle  \langle \tilde{a}^\dagger  \tilde{a}_0\rangle  -2 \langle \tilde{a}^\dagger\rangle  \langle \tilde{a}_0\rangle  \langle \tilde\sigma^{1}_{12}\rangle  \right) e^{ i {\phi_{2}(t)}} \\
&- i g \left( \langle \tilde{a}\rangle  \langle \tilde\sigma^{1}_{21}  \tilde{a}_0\rangle  + \langle \tilde{a}_0\rangle  \langle \tilde{a}  \tilde\sigma^{1}_{21}\rangle  + \langle \tilde\sigma^{1}_{21}\rangle  \langle \tilde{a}  \tilde{a}_0\rangle  -2 \langle \tilde{a}\rangle  \langle \tilde{a}_0\rangle  \langle \tilde\sigma^{1}_{21}\rangle  \right) e^{ i {\phi_{2}(t)}} - \Gamma_{2} \langle \tilde\sigma^{1}_{22}  \tilde{a}_0\rangle  e^{ i {\phi_{2}(t)}} \\
&- i \Omega_{2} \langle \tilde\sigma^{1}_{21}  \tilde{a}_0\rangle +  i \Omega_{2} \langle \tilde\sigma^{1}_{12}  \tilde{a}_0\rangle e^{2 i {\phi_{2}(t)}}
\end{aligned}
\label{eq:corr_sa_1}
\end{equation}
\begin{equation}
\begin{aligned}
\frac{d}{dt} \langle \tilde\sigma^{1}_{12}  \tilde{a}_0\rangle e^{2 i {\phi_{2}(t)}} &= 2 i \dot\phi_{2} \langle \tilde\sigma^{1}_{12}  \tilde{a}_0\rangle  + 2 i g \left( \langle \tilde{a}\rangle  \langle \tilde\sigma^{1}_{22}  \tilde{a}_0\rangle  + \langle \tilde{a}_0\rangle  \langle \tilde{a}  \tilde\sigma^{1}_{22}\rangle  + \langle \tilde\sigma^{1}_{22}\rangle  \langle \tilde{a}  \tilde{a}_0\rangle  -2 \langle \tilde{a}\rangle  \langle \tilde{a}_0\rangle  \langle \tilde\sigma^{1}_{22}\rangle  \right) e^{2 i {\phi_{2}(t)}} \\
&- i \Omega_{2} \langle \tilde{a}_0\rangle e^{ i {\phi_{2}(t)}} - i g \langle \tilde{a}  \tilde{a}_0\rangle  e^{2 i {\phi_{2}(t)}} -0.5 \Gamma_{2} \langle \tilde\sigma^{1}_{12}  \tilde{a}_0\rangle  e^{2 i {\phi_{2}(t)}} +  i \Delta_{2} \langle \tilde\sigma^{1}_{12}  \tilde{a}_0\rangle  e^{2 i {\phi_{2}(t)}} \\
&+ 2 i \Omega_{2} \langle \tilde\sigma^{1}_{22}  \tilde{a}_0\rangle  e^{ i {\phi_{2}(t)}}
\end{aligned}
\label{eq:corr_sa_2}
\end{equation}
\label{eq:corr_sa}
\end{subequations}
In equations \eqref{eq:corr_sa_1} and \eqref{eq:corr_sa_2}, terms proportional to $\dot{\phi}$ appear, namely $i \dot{\phi} \braket{\tilde{\sigma}_{22}^1 \tilde{a}_0}$ and $2 i \dot{\phi} \braket{\tilde{\sigma}_{12}^1 \tilde{a}_0}$, respectively. Transforming these Stratonovich stochastic differential equations to It\^o as in section \ref{sec:dephasing_pump} leads to a dephasing term $- 0.5 \nu_2 \braket{\tilde{\sigma}_{22}^1 \tilde{a}_0}$ and $- 2 \nu_2 \braket{\tilde{\sigma}_{22}^1 \tilde{a}_0}$, respectively. In section \ref{sec:corr_fct_eqs} you find the closed system of differential equations derived via this procedure.

\newpage
\section{Second-order cumulant expansion}

In order to derive the following equations we wrote a program~\cite{github_qumulants} that symbolically evaluates bosonic and fermionic commutation relations. Then, the generalized cumulant expansion is applied to obtain a closed set of equations featuring only first- and second-order averages. The correctness of the equations is ensured by comparing numerical results for smaller systems with a full quantum treatment for a variety of parameters.

\subsection{System equations} \label{sec:sys_eqs}
Note, that we only show part of the derived equations in order to keep the length of the entire set at a comprehensible level. As previously mentioned, the (in total 37) equations were derived using a software tool. The tool is open source and available online~\cite{github_qumulants}.
\begin{flalign*}\begin{autobreak}
\frac{d}{dt} \langle a^\dagger a\rangle =
 - \kappa \langle a^\dagger a\rangle
 + i N g \langle a \sigma^{1}_{21}\rangle
 -i N g \langle a^\dagger \sigma^{1}_{12}\rangle 
\end{autobreak}&\end{flalign*}\vspace{-0.7cm}\begin{flalign*}\begin{autobreak}\frac{d}{dt} \langle a\rangle =
 i \Delta_{c} \langle a\rangle -0.5 \left( \kappa + \nu_{2} \right) \langle a\rangle
 -i N g \langle \sigma^{1}_{12}\rangle 
\end{autobreak}&\end{flalign*}\vspace{-0.7cm}\begin{flalign*}\begin{autobreak}\frac{d}{dt} \langle \sigma^{1}_{22}\rangle =
 - \Gamma_{2} \langle \sigma^{1}_{22}\rangle
 + i \Omega_{2} \langle \sigma^{1}_{12}\rangle
 + i g \langle a^\dagger \sigma^{1}_{12}\rangle
 -i \Omega_{2} \langle \sigma^{1}_{21}\rangle
 -i g \langle a \sigma^{1}_{21}\rangle 
\end{autobreak}&\end{flalign*}\vspace{-0.7cm}\begin{flalign*}\begin{autobreak}\frac{d}{dt} \langle \sigma^{1}_{12}\rangle =
 i \Delta_{2} \langle \sigma^{1}_{12}\rangle -0.5 \left( \Gamma_{2} + \nu_{2} \right) \langle \sigma^{1}_{12}\rangle
 + i \Omega_{3} \langle \sigma^{1}_{32}\rangle
 + i \Omega_{2} \left( -1 + \langle \sigma^{1}_{33}\rangle \right)
 + i g \langle a \sigma^{1}_{33}\rangle
 -i g \langle a\rangle
 + 2 i  \Omega_{2} \langle \sigma^{1}_{22}\rangle
 + 2 i  g \langle a \sigma^{1}_{22}\rangle 
\end{autobreak}&\end{flalign*}\vspace{-0.7cm}\begin{flalign*}\begin{autobreak}\frac{d}{dt} \langle \sigma^{1}_{32}\rangle =
 -0.5 \left( \Gamma_{2} + \Gamma_{3} + \nu_{2} + \nu_{3} \right) \langle \sigma^{1}_{32}\rangle
 + i \Omega_{3} \langle \sigma^{1}_{12}\rangle
 + i \Delta_{2} \langle \sigma^{1}_{32}\rangle
 -i \Omega_{2} \langle \sigma^{1}_{31}\rangle
 -i \Delta_{3} \langle \sigma^{1}_{32}\rangle
 -i g \langle a \sigma^{1}_{31}\rangle 
\end{autobreak}&\end{flalign*}\vspace{-0.7cm}\begin{flalign*}\begin{autobreak}\frac{d}{dt} \langle \sigma^{1}_{33}\rangle =
 - \Gamma_{3} \langle \sigma^{1}_{33}\rangle
 + i \Omega_{3} \langle \sigma^{1}_{13}\rangle
 -i \Omega_{3} \langle \sigma^{1}_{31}\rangle 
\end{autobreak}&\end{flalign*}\vspace{-0.7cm}\begin{flalign*}\begin{autobreak}\frac{d}{dt} \langle a \sigma^{1}_{21}\rangle =
 i \Omega_{2} \langle a\rangle
 + i g \langle a^\dagger a\rangle
 + i \Delta_{c} \langle a \sigma^{1}_{21}\rangle -0.5 \Gamma_{2} \langle a \sigma^{1}_{21}\rangle -0.5 \kappa \langle a \sigma^{1}_{21}\rangle
 -i g \langle \sigma^{1}_{22}\rangle
 -2 i  g \left( \langle a^\dagger\rangle \langle a \sigma^{1}_{22}\rangle + \langle a\rangle \langle a^\dagger \sigma^{1}_{22}\rangle + \langle \sigma^{1}_{22}\rangle \langle a^\dagger a\rangle -2 \langle a^\dagger\rangle \langle a\rangle \langle \sigma^{1}_{22}\rangle \right)
 -i g \left( \langle a^\dagger\rangle \langle a \sigma^{1}_{33}\rangle + \langle a\rangle \langle a^\dagger \sigma^{1}_{33}\rangle + \langle \sigma^{1}_{33}\rangle \langle a^\dagger a\rangle -2 \langle a^\dagger\rangle \langle a\rangle \langle \sigma^{1}_{33}\rangle \right)
 -i \Delta_{2} \langle a \sigma^{1}_{21}\rangle
 -2 i  \Omega_{2} \langle a \sigma^{1}_{22}\rangle
 -i \Omega_{2} \langle a \sigma^{1}_{33}\rangle
 -i \Omega_{3} \langle a \sigma^{1}_{23}\rangle
 -i g \left( -1 + N \right) \langle \sigma^{1}_{21} \sigma^{2}_{12}\rangle 
\end{autobreak}&\end{flalign*}\vspace{-0.7cm}\begin{flalign*}\begin{autobreak}\frac{d}{dt} \langle a \sigma^{1}_{22}\rangle =
 i g \left( \langle a^\dagger\rangle \langle a \sigma^{1}_{12}\rangle + \langle a\rangle \langle a^\dagger \sigma^{1}_{12}\rangle + \langle \sigma^{1}_{12}\rangle \langle a^\dagger a\rangle -2 \langle a^\dagger\rangle \langle a\rangle \langle \sigma^{1}_{12}\rangle \right)
 + i \Delta_{c} \langle a \sigma^{1}_{22}\rangle
 + i \Omega_{2} \langle a \sigma^{1}_{12}\rangle - \Gamma_{2} \langle a \sigma^{1}_{22}\rangle -0.5 \kappa \langle a \sigma^{1}_{22}\rangle -0.5 \nu_{2} \langle a \sigma^{1}_{22}\rangle
 -i g \left( \langle \sigma^{1}_{21}\rangle \langle a a\rangle + 2 \langle a\rangle \langle a \sigma^{1}_{21}\rangle -2 \langle \sigma^{1}_{21}\rangle \langle a\rangle ^{2} \right)
 -i \Omega_{2} \langle a \sigma^{1}_{21}\rangle
 -i g \left( -1 + N \right) \langle \sigma^{1}_{22} \sigma^{2}_{12}\rangle 
\end{autobreak}&\end{flalign*}\vspace{-0.7cm}\begin{flalign*}\begin{autobreak}\frac{d}{dt} \langle \sigma^{1}_{21} \sigma^{2}_{12}\rangle =
 i g \left( \langle a\rangle \langle \sigma^{1}_{33} \sigma^{2}_{21}\rangle + \langle \sigma^{1}_{33}\rangle \langle a \sigma^{1}_{21}\rangle + \langle \sigma^{1}_{21}\rangle \langle a \sigma^{1}_{33}\rangle -2 \langle a\rangle \langle \sigma^{1}_{33}\rangle \langle \sigma^{1}_{21}\rangle \right)
 + i \Omega_{2} \langle \sigma^{1}_{12}\rangle
 + i g \langle a^\dagger \sigma^{1}_{12}\rangle
 + i \Omega_{2} \langle \sigma^{1}_{33} \sigma^{2}_{21}\rangle
 + i \Omega_{3} \langle \sigma^{1}_{21} \sigma^{2}_{32}\rangle - \Gamma_{2} \langle \sigma^{1}_{21} \sigma^{2}_{12}\rangle
 -2 i  g \left( \langle a^\dagger\rangle \langle \sigma^{1}_{22} \sigma^{2}_{12}\rangle + \langle \sigma^{1}_{22}\rangle \langle a^\dagger \sigma^{1}_{12}\rangle + \langle \sigma^{1}_{12}\rangle \langle a^\dagger \sigma^{1}_{22}\rangle -2 \langle a^\dagger\rangle \langle \sigma^{1}_{22}\rangle \langle \sigma^{1}_{12}\rangle \right)
 -i g \left( \langle a^\dagger\rangle \langle \sigma^{1}_{33} \sigma^{2}_{12}\rangle + \langle \sigma^{1}_{33}\rangle \langle a^\dagger \sigma^{1}_{12}\rangle + \langle \sigma^{1}_{12}\rangle \langle a^\dagger \sigma^{1}_{33}\rangle -2 \langle a^\dagger\rangle \langle \sigma^{1}_{33}\rangle \langle \sigma^{1}_{12}\rangle \right)
 + 2 i  g \left( \langle a\rangle \langle \sigma^{1}_{22} \sigma^{2}_{21}\rangle + \langle \sigma^{1}_{22}\rangle \langle a \sigma^{1}_{21}\rangle + \langle \sigma^{1}_{21}\rangle \langle a \sigma^{1}_{22}\rangle -2 \langle a\rangle \langle \sigma^{1}_{22}\rangle \langle \sigma^{1}_{21}\rangle \right)
 -i \Omega_{2} \langle \sigma^{1}_{21}\rangle
 -i g \langle a \sigma^{1}_{21}\rangle
 -2 i  \Omega_{2} \langle \sigma^{1}_{22} \sigma^{2}_{12}\rangle
 + 2 i  \Omega_{2} \langle \sigma^{1}_{22} \sigma^{2}_{21}\rangle
 -i \Omega_{2} \langle \sigma^{1}_{33} \sigma^{2}_{12}\rangle
 -i \Omega_{3} \langle \sigma^{1}_{12} \sigma^{2}_{23}\rangle 
\end{autobreak}&\end{flalign*}\vspace{-0.7cm}\begin{flalign*}\begin{autobreak}\frac{d}{dt} \langle \sigma^{1}_{22} \sigma^{2}_{22}\rangle =
 i g \left( 2 \langle a^\dagger\rangle \langle \sigma^{1}_{22} \sigma^{2}_{12}\rangle + 2 \langle \sigma^{1}_{22}\rangle \langle a^\dagger \sigma^{1}_{12}\rangle + 2 \langle \sigma^{1}_{12}\rangle \langle a^\dagger \sigma^{1}_{22}\rangle -4 \langle a^\dagger\rangle \langle \sigma^{1}_{22}\rangle \langle \sigma^{1}_{12}\rangle \right) -2 \Gamma_{2} \langle \sigma^{1}_{22} \sigma^{2}_{22}\rangle
 -2 i  g \left( \langle a\rangle \langle \sigma^{1}_{22} \sigma^{2}_{21}\rangle + \langle \sigma^{1}_{22}\rangle \langle a \sigma^{1}_{21}\rangle + \langle \sigma^{1}_{21}\rangle \langle a \sigma^{1}_{22}\rangle -2 \langle a\rangle \langle \sigma^{1}_{22}\rangle \langle \sigma^{1}_{21}\rangle \right)
 + 2 i  \Omega_{2} \langle \sigma^{1}_{22} \sigma^{2}_{12}\rangle
 -2 i  \Omega_{2} \langle \sigma^{1}_{22} \sigma^{2}_{21}\rangle 
\end{autobreak}&\end{flalign*}

\newpage
\subsection{Correlation function equations} \label{sec:corr_fct_eqs}
We omit the phase factors in these equations, since this only corresponds to a variable relabeling of some specific correlation functions. 
\begin{flalign*}\begin{autobreak}
\frac{d}{dt} \langle \tilde{a}^\dagger \tilde{a}_0\rangle =

 - i \Delta_{c} \langle \tilde{a}^\dagger \tilde{a}_0\rangle -0.5 \kappa \langle \tilde{a}^\dagger \tilde{a}_0\rangle
 + i N g \langle \tilde{\sigma}^{1}_{21} \tilde{a}_0\rangle 
\end{autobreak}&\end{flalign*}\vspace{-0.6cm}\begin{flalign*}\begin{autobreak}\frac{d}{dt} \langle \tilde{\sigma}^{1}_{21} \tilde{a}_0\rangle =
 -2 i g \left( \langle a^\dagger\rangle \langle \tilde{\sigma}^{1}_{22} \tilde{a}_0\rangle + \langle a\rangle \langle a^\dagger \sigma^{1}_{22}\rangle + \langle \sigma^{1}_{22}\rangle \langle \tilde{a}^\dagger \tilde{a}_0\rangle -2 \langle a^\dagger\rangle \langle a\rangle \langle \sigma^{1}_{22}\rangle \right)
 - i g \left( \langle a^\dagger\rangle \langle \tilde{\sigma}^{1}_{33} \tilde{a}_0\rangle + \langle a\rangle \langle a^\dagger \sigma^{1}_{33}\rangle + \langle \sigma^{1}_{33}\rangle \langle \tilde{a}^\dagger \tilde{a}_0\rangle -2 \langle a^\dagger\rangle \langle a\rangle \langle \sigma^{1}_{33}\rangle \right)
 + i \Omega_{2} \langle a\rangle
 + i g \langle \tilde{a}^\dagger \tilde{a}_0\rangle -0.5 \Gamma_{2} \langle \tilde{\sigma}^{1}_{21} \tilde{a}_0\rangle
 - i \Delta_{2} \langle \tilde{\sigma}^{1}_{21} \tilde{a}_0\rangle -2 i \Omega_{2} \langle \tilde{\sigma}^{1}_{22} \tilde{a}_0\rangle
 - i \Omega_{2} \langle \tilde{\sigma}^{1}_{33} \tilde{a}_0\rangle
 - i \Omega_{3} \langle \tilde{\sigma}^{1}_{23} \tilde{a}_0\rangle 
\end{autobreak}&\end{flalign*}\vspace{-0.6cm}\begin{flalign*}\begin{autobreak}\frac{d}{dt} \langle \tilde{\sigma}^{1}_{31} \tilde{a}_0\rangle =

 - i g \left( \langle a^\dagger\rangle \langle \tilde{\sigma}^{1}_{32} \tilde{a}_0\rangle + \langle a\rangle \langle a^\dagger \sigma^{1}_{32}\rangle + \langle \sigma^{1}_{32}\rangle \langle \tilde{a}^\dagger \tilde{a}_0\rangle -2 \langle a^\dagger\rangle \langle a\rangle \langle \sigma^{1}_{32}\rangle \right)
 + i \Omega_{3} \langle a\rangle -0.5 \Gamma_{3} \langle \tilde{\sigma}^{1}_{31} \tilde{a}_0\rangle
 - i \Delta_{3} \langle \tilde{\sigma}^{1}_{31} \tilde{a}_0\rangle
 - i \Omega_{2} \langle \tilde{\sigma}^{1}_{32} \tilde{a}_0\rangle
 - i \Omega_{3} \langle \tilde{\sigma}^{1}_{22} \tilde{a}_0\rangle -2 i \Omega_{3} \langle \tilde{\sigma}^{1}_{33} \tilde{a}_0\rangle -0.5 \gamma_{2} \langle \tilde{\sigma}^{1}_{31} \tilde{a}_0\rangle -0.5 \gamma_{3} \langle \tilde{\sigma}^{1}_{31} \tilde{a}_0\rangle 
\end{autobreak}&\end{flalign*}\vspace{-0.6cm}\begin{flalign*}\begin{autobreak}\frac{d}{dt} \langle \tilde{\sigma}^{1}_{22} \tilde{a}_0\rangle =

 - i g \left( \langle a\rangle \left( \langle a \sigma^{1}_{21}\rangle + \langle \tilde{\sigma}^{1}_{21} \tilde{a}_0\rangle \right) + \langle \sigma^{1}_{21}\rangle \langle \tilde{a} \tilde{a}_0\rangle -2 \langle \sigma^{1}_{21}\rangle \langle a\rangle ^{2} \right)
 + i g \left( \langle a^\dagger\rangle \langle \tilde{\sigma}^{1}_{12} \tilde{a}_0\rangle + \langle a\rangle \langle a^\dagger \sigma^{1}_{12}\rangle + \langle \sigma^{1}_{12}\rangle \langle \tilde{a}^\dagger \tilde{a}_0\rangle -2 \langle a^\dagger\rangle \langle a\rangle \langle \sigma^{1}_{12}\rangle \right) - \Gamma_{2} \langle \tilde{\sigma}^{1}_{22} \tilde{a}_0\rangle
 - i \Omega_{2} \langle \tilde{\sigma}^{1}_{21} \tilde{a}_0\rangle
 + i \Omega_{2} \langle \tilde{\sigma}^{1}_{12} \tilde{a}_0\rangle -0.5 \gamma_{2} \langle \tilde{\sigma}^{1}_{22} \tilde{a}_0\rangle 
\end{autobreak}&\end{flalign*}\vspace{-0.6cm}\begin{flalign*}\begin{autobreak}\frac{d}{dt} \langle \tilde{\sigma}^{1}_{32} \tilde{a}_0\rangle =

 - i g \left( \langle a\rangle \left( \langle a \sigma^{1}_{31}\rangle + \langle \tilde{\sigma}^{1}_{31} \tilde{a}_0\rangle \right) + \langle \sigma^{1}_{31}\rangle \langle \tilde{a} \tilde{a}_0\rangle -2 \langle \sigma^{1}_{31}\rangle \langle a\rangle ^{2} \right) -0.5 \Gamma_{2} \langle \tilde{\sigma}^{1}_{32} \tilde{a}_0\rangle -0.5 \Gamma_{3} \langle \tilde{\sigma}^{1}_{32} \tilde{a}_0\rangle
 + i \Delta_{2} \langle \tilde{\sigma}^{1}_{32} \tilde{a}_0\rangle
 - i \Delta_{3} \langle \tilde{\sigma}^{1}_{32} \tilde{a}_0\rangle
 - i \Omega_{2} \langle \tilde{\sigma}^{1}_{31} \tilde{a}_0\rangle
 + i \Omega_{3} \langle \tilde{\sigma}^{1}_{12} \tilde{a}_0\rangle -2 \gamma_{2} \langle \tilde{\sigma}^{1}_{32} \tilde{a}_0\rangle -0.5 \gamma_{3} \langle \tilde{\sigma}^{1}_{32} \tilde{a}_0\rangle 
\end{autobreak}&\end{flalign*}\vspace{-0.6cm}\begin{flalign*}\begin{autobreak}\frac{d}{dt} \langle \tilde{\sigma}^{1}_{33} \tilde{a}_0\rangle =
 - \Gamma_{3} \langle \tilde{\sigma}^{1}_{33} \tilde{a}_0\rangle
 - i \Omega_{3} \langle \tilde{\sigma}^{1}_{31} \tilde{a}_0\rangle
 + i \Omega_{3} \langle \tilde{\sigma}^{1}_{13} \tilde{a}_0\rangle -0.5 \gamma_{2} \langle \tilde{\sigma}^{1}_{33} \tilde{a}_0\rangle 
\end{autobreak}&\end{flalign*}\vspace{-0.6cm}\begin{flalign*}\begin{autobreak}\frac{d}{dt} \langle \tilde{a} \tilde{a}_0\rangle =
 i \Delta_{c} \langle \tilde{a} \tilde{a}_0\rangle -2 \gamma_{2} \langle \tilde{a} \tilde{a}_0\rangle -0.5 \kappa \langle \tilde{a} \tilde{a}_0\rangle
 - i N g \langle \tilde{\sigma}^{1}_{12} \tilde{a}_0\rangle 
\end{autobreak}&\end{flalign*}\vspace{-0.6cm}\begin{flalign*}\begin{autobreak}\frac{d}{dt} \langle \tilde{\sigma}^{1}_{12} \tilde{a}_0\rangle =
 2 i g \left( \langle a\rangle \left( \langle a \sigma^{1}_{22}\rangle + \langle \tilde{\sigma}^{1}_{22} \tilde{a}_0\rangle \right) + \langle \sigma^{1}_{22}\rangle \langle \tilde{a} \tilde{a}_0\rangle -2 \langle \sigma^{1}_{22}\rangle \langle a\rangle ^{2} \right)
 + i g \left( \langle a\rangle \left( \langle a \sigma^{1}_{33}\rangle + \langle \tilde{\sigma}^{1}_{33} \tilde{a}_0\rangle \right) + \langle \sigma^{1}_{33}\rangle \langle \tilde{a} \tilde{a}_0\rangle -2 \langle \sigma^{1}_{33}\rangle \langle a\rangle ^{2} \right)
 - i \Omega_{2} \langle a\rangle
 - i g \langle \tilde{a} \tilde{a}_0\rangle -0.5 \Gamma_{2} \langle \tilde{\sigma}^{1}_{12} \tilde{a}_0\rangle
 + i \Delta_{2} \langle \tilde{\sigma}^{1}_{12} \tilde{a}_0\rangle + 2 i \Omega_{2} \langle \tilde{\sigma}^{1}_{22} \tilde{a}_0\rangle
 + i \Omega_{2} \langle \tilde{\sigma}^{1}_{33} \tilde{a}_0\rangle
 + i \Omega_{3} \langle \tilde{\sigma}^{1}_{32} \tilde{a}_0\rangle -2 \gamma_{2} \langle \tilde{\sigma}^{1}_{12} \tilde{a}_0\rangle 
\end{autobreak}&\end{flalign*}\vspace{-0.6cm}\begin{flalign*}\begin{autobreak}\frac{d}{dt} \langle \tilde{\sigma}^{1}_{13} \tilde{a}_0\rangle =
 i g \left( \langle a\rangle \left( \langle a \sigma^{1}_{23}\rangle + \langle \tilde{\sigma}^{1}_{23} \tilde{a}_0\rangle \right) + \langle \sigma^{1}_{23}\rangle \langle \tilde{a} \tilde{a}_0\rangle -2 \langle \sigma^{1}_{23}\rangle \langle a\rangle ^{2} \right)
 - i \Omega_{3} \langle a\rangle -0.5 \Gamma_{3} \langle \tilde{\sigma}^{1}_{13} \tilde{a}_0\rangle
 + i \Delta_{3} \langle \tilde{\sigma}^{1}_{13} \tilde{a}_0\rangle
 + i \Omega_{2} \langle \tilde{\sigma}^{1}_{23} \tilde{a}_0\rangle
 + i \Omega_{3} \langle \tilde{\sigma}^{1}_{22} \tilde{a}_0\rangle + 2 i \Omega_{3} \langle \tilde{\sigma}^{1}_{33} \tilde{a}_0\rangle -0.5 \gamma_{2} \langle \tilde{\sigma}^{1}_{13} \tilde{a}_0\rangle -0.5 \gamma_{3} \langle \tilde{\sigma}^{1}_{13} \tilde{a}_0\rangle 
\end{autobreak}&\end{flalign*}\vspace{-0.6cm}\begin{flalign*}\begin{autobreak}\frac{d}{dt} \langle \tilde{\sigma}^{1}_{23} \tilde{a}_0\rangle =
 i g \left( \langle a^\dagger\rangle \langle \tilde{\sigma}^{1}_{13} \tilde{a}_0\rangle + \langle a\rangle \langle a^\dagger \sigma^{1}_{13}\rangle + \langle \sigma^{1}_{13}\rangle \langle \tilde{a}^\dagger \tilde{a}_0\rangle -2 \langle a^\dagger\rangle \langle a\rangle \langle \sigma^{1}_{13}\rangle \right) -0.5 \Gamma_{2} \langle \tilde{\sigma}^{1}_{23} \tilde{a}_0\rangle -0.5 \Gamma_{3} \langle \tilde{\sigma}^{1}_{23} \tilde{a}_0\rangle
 - i \Delta_{2} \langle \tilde{\sigma}^{1}_{23} \tilde{a}_0\rangle
 + i \Delta_{3} \langle \tilde{\sigma}^{1}_{23} \tilde{a}_0\rangle
 + i \Omega_{2} \langle \tilde{\sigma}^{1}_{13} \tilde{a}_0\rangle
 - i \Omega_{3} \langle \tilde{\sigma}^{1}_{21} \tilde{a}_0\rangle -0.5 \gamma_{3} \langle \tilde{\sigma}^{1}_{23} \tilde{a}_0\rangle 
\end{autobreak}&\end{flalign*}


\end{document}